\documentclass[12pt,a4paper]{article}
\usepackage{jheppub}
\usepackage{hyperref}

\usepackage{mathtext}         % ?? ?? ?????? ????? (? ?????)

\usepackage{amsmath,amssymb} 
\usepackage{amsfonts}
\usepackage{epsfig}

\newcommand{\scr}{\scriptscriptstyle}

\newcommand{\be}{\begin{equation}}
\newcommand{\ee}{\end{equation}}
\newcommand{\bea}{\begin{eqnarray}}
\newcommand{\eea}{\end{eqnarray}}
\newcommand{\ba}{\begin{eqnarray*}}
\newcommand{\ea}{\end{eqnarray*}}

\newcommand{\FmS}[2]{\Fs{#1}{#2}{\frac{4\widetilde{m}^2}{q^2}}}
\newcommand{\FCP}[2]{\Fs{#1}{#2}{1-\frac{\widetilde{\mu}_{3}}{\widetilde{\mu}_4}}}
\newcommand{\FMU}[2]{\Fs{#1}{#2}{\frac{\widetilde{\mu}_3}{\widetilde{\mu}_4}}}
\newcommand{\FMM}[2]{\Fs{#1}{#2}{\frac{\widetilde{\mu}_3}{\widetilde{\mu}_4}}}
\newcommand{\Fe}[2]{\Fs{#1}{#2}{z}}
\newcommand{\Fh}[2]{\,{}_#1F_#2}
\newcommand{\Fs}[3]{\!\!\left[\begin{array}{c}#1\,;\\#2\,;\end{array}#3\right]}
\newcommand{\Fsm}[2]{\Fs{#1}{#2}{\frac{-\widetilde{s}_{ij}}{4\widetilde{\mu}_4}}}
\newcommand{\Fx}[2]{\Fs{#1}{#2}{x}}
\newcommand{\Fy}[2]{\Fs{#1}{#2}{y}}
\newcommand{\Ffp}[2]{\Fs{#1}{#2}{\frac{ q^2}{4m^2}}}
\newcommand{\Ffw}[2]{\Fs{#1}{#2}{\frac{-\widetilde{q}^{\:2}}{4\widetilde{m}^2}}}

\begin{document}

\begin{titlepage}

\begin{flushright}
\today  \\
\date \\
\end{flushright}

\vspace*{0.2cm}
\begin{center}
{\Large {Functional reduction of Feynman integrals
}}\\[2 cm]
\end{center}

\begin{center}
{\bf  O.V.~Tarasov}
\vspace{0.5cm}
\\   

\it Joint Institute for Nuclear Research,\\
      141980 Dubna, Russian Federation \\
     E-mail: {\tt otarasov@jinr.ru}

\end{center}

\vspace*{1.0cm}

\begin{abstract}

 A method for reducing  Feynman integrals, depending on 
several kinematic variables and masses, to a combination of
integrals with fewer  variables is proposed.
The method is based on iterative application of functional
equations proposed by the author. The reduction of the one-loop scalar triangle  
and box  integrals with 
massless internal propagators to  simpler integrals is described in detail.
The triangle integral  depending on three variables 
is represented as a sum over three integrals depending on two variables.  
By solving the dimensional recurrence relations 
for these integrals, an analytic expression in terms of  
the $_2F_1$ Gauss hypergeometric
function   and the logarithmic function was derived.

By using the functional equations, the one-loop box integral with 
massless internal propagators, which depends on six kinematic variables, 
was expressed   as a sum of 12 terms. These  terms are proportional 
to the same  integral depending only on three   variables
different for each term.  
For this integral with three variables, an analytic result
in terms of the  $F_1$ Appell and $_2F_1$ Gauss  hypergeometric 
functions was derived by solving  the recurrence  relation with respect
to the spacetime dimension $d$. 
The reduction equations for the box integral with  some  
kinematic variables equal to zero are considered.

\end{abstract}

\end{titlepage}
\tableofcontents

\section{Introduction}

%%%%%%%%%%%%%%%%%%%%%%%%%%%%%%%%%%%%%%%%%%%%%%%%%%%%%%%%%%%%%%%%%
%%%%%%%%%%%%%%%%%%%%%%%%%%%%%%%%%%%%%%%%%%%%%%%%%%%%%%%%%%%%%%%%
Theoretical predictions for  experiments at the LHC
\cite{Aad:2012tfa,Chatrchyan:2012xdj} as well as
at future colliders such as the FCC \cite{Mangano:2651294}
demand  knowledge of  precise
radiative corrections. Precise experimental measurements
are to be interpreted with  sufficient precision of theoretical 
predictions.
The problem of calculating  such radiative corrections
is associated, in particular, with the need to compute 
Feynman integrals depending on several kinematic
variables and/or masses.
Over the past few decades
enormous progress has been made in solving the
problem of evaluating Feynman integrals.

However, 
for further progress in the analytic evaluation of Feynman integrals,
especially integrals depending on several kinematic variables and masses,
new mathematical methods need to be elaborated. In this respect,
more and more attention is paid to  methods  based on solving
different kinds of recurrence relations.
The first result obtained in this approach was  analytic 
evaluation of the two-loop propagator integral with massless 
propagators   \cite{Kazakov:1983pk}. In this paper, an analytic
result for the integral was derived by solving the recurrence relation
with respect to the power of a propagator.
A systematic approach for evaluating  Feynman integrals
by solving the recurrence relation with respect to the power
of a propagator  was described in Ref.\cite{Laporta:2001dd}.

In Ref. \cite{Tarasov:1996br}, a method 
for evaluating Feynman integrals
based on the recurrence relations with respect
to the space-time  dimension $d$ was suggested.
It turns out that hypergeometric functions appearing 
in the solution of the recurrence relations
with respect to $d$ \cite{Tarasov:2000sf},\cite{Fleischer:2003rm} 
have fewer arguments than in the results for these integrals
obtained by other methods. For example, the results obtained
in Refs.  \cite{Davydychev:1990jt}, \cite{Davydychev:1990cq}
using the Mellin-Barnes integration technique,
and in Ref. \cite{Anastasiou:1999ui} using the negative dimension method 
are expressed in terms of  hypergeometric functions with more arguments than
those obtained for these integrals by solving dimensional recurrence relations.
\\

In Ref.\cite{Tarasov:2008hw},  new relationships
between the Feynman integrals with different kinematic
variables were discovered.
A method of deriving   functional equations
from algebraic relations for products of propagators
was recently proposed in Ref.\cite{Tarasov:2017}.
At the one-loop level some functional relationships were
also considered in Refs. \cite{Davydychev:2016dfi},
\cite{Davydychev:2017bbl}.

It was shown that these relationships, or in other words
functional equations, can be used
to express Feynman integrals in terms of integrals
with fewer variables. In Ref. \cite{Kniehl:2009pv},   
the functional equations were used to obtain 
relations between integrals appearing in 
 radiative corrections for different physical processes.
\\
An 
important step in evaluating radiative corrections for physical
processes  is the Laurent expansion in the $\varepsilon=(4-d)/2$ 
of the analytic results for  Feynman integrals.
Quite essential progress in this field was made in many papers.
See, for example,\cite{Davydychev:1999mq,Davydychev:2000kw, 
Davydychev:2000na, Huber:2005yg,Kalmykov:2006hu,Kalmykov:2006pu}.
Up to now the Laurent expansion of Feynman integrals in
 $\varepsilon$  is not a completely solved problem.
Even at the one-loop level  only the first several terms in
the $\varepsilon$ expansion of the four- and higher point 
functions are known. 
The existing results \cite{Nierste:1992wg}, \cite{Korner:2004rr} 
are not so easy to generalize for integrals depending on several
masses and/or several external off-shell momenta.

It is evident that for the $\varepsilon$ expansion 
the simplicity of the analytical results for dimensionally 
regularized integrals is rather important. 
For this reason
the method based on the solution of recurrence relations 
and the method of functional reduction suggested
in this article can be very useful.

In the present  paper, we propose a framework for systematically 
 reducing Feynman integrals depending on several
kinematic variables and masses  to a combination of integrals with 
fewer variables.  This framework is based on solving the functional 
equations for  Feynman  integrals proposed in Refs. 
\cite{Tarasov:2008hw,Tarasov:2011zz,Tarasov:2017}.
The main steps of our approach will be illustrated on the  one-loop 
integrals with massless propagators. 

In a sense, the application of functional equations 
for evaluating integrals is analogous to the   use of
recurrence relations with respect to some discretely
changing parameters, like space-time dimension $d$ or power
of a propagator. Applying such recurrence relations, 
one can reduce an integral to a set of basis integrals
which are in fact boundary values of the integrals
of interest.
Using functional equations one can reduce an integral
to a combination of integrals with fewer variables, i.e. integrals
defined on some hypersurfaces. In other words, these integrals can be
interpreted  as a kind of boundary integrals. 
\\

This paper is organized as follows.

In section 2, we briefly discuss the method of discovering
functional equations for Feynman integrals and describe the methods
of obtaining their solutions.
As an illustrative example, we  consider the solution
of the functional equation for the one-loop propagator
integral with arbitrary masses.
\\

In section 3, the one-loop integral associated with the triangle 
Feynman diagram with massless internal propagators is considered.
We present the functional equation for this integral and describe
its solution. The analytic result for the integral
appearing in the solution of thefunctional equation is obtained
as a solution of the dimensional recurrence relation.
A particular case of the functional equation
for the triangle integral is considered. 
\\ 

In section 4, we present the functional equations for the 
one-loop scalar  integral associated with the Feynman diagram with four
external legs. A two step procedure, based on  functional equations,
for reducing  the integral 
depending on six variables to a combination of integrals
depending on three variables is described.
For these integrals, depending on three variables,
an analytic result as a solution of the dimensional recurrence
relation is presented.
 Functional reduction of the box integral for several
particular cases of kinematic variables is considered.
The first few terms in the Laurent expansion around $d=4$ and $d=6$
for these integrals are given.
\\

In section 5, we report our conclusions and discuss 
future applications of functional equations for evaluating
Feynman integrals corresponding to diagrams with massive internal lines 
and  with more external legs and loops.
\\

In Appendix A, we present definitions and explicit formulae for the
Gram determinants and polynomials occurring in the paper.
In appendix B, a derivation of the analytic result for the one-loop integral with
massless internal  propagators  with  particular emphasis on its dependence
on the small imaginary part needed for the correct analytic continuation
of the integral is presented.
In Appendix C, the series and integral representations for the
hypergeometric functions used in the paper are given.

%%%%%%%%%%%%%%%%%%%%%%%%%%%%%%%%%%%%%%%%%%%%%%%%%%%%%%%
\section{Functional equations and their solution}
%%%%%%%%%%%%%%%%%%%%%%%%%%%%%%%%%%%%%%%%%%%%%%%%%%%%%%%

At the present time, there are three methods for deriving functional
equations for Feynman integrals. The method proposed in Ref.
\cite{Tarasov:2008hw} is based on exploiting recurrence relations
obtained by the method of generalized recurrence relations
\cite{Tarasov:1996br}. By choosing some kinematic variables, one 
can eliminate most complicated integrals from the recurrence relation
so that the sum of remaining terms represents the functional equation.
The second method is based on  algebraic relations for a sum of
products of propagators \cite{Tarasov:2017}. Integrating such sums 
with respect to a common to all propagators momentum
 gives a functional equation.
The third method is based on the use of algebraic relations for modified propagators
\cite{Tarasov:2017}.
Integrating an algebraic relation depending on  modified propagators with 
respect to a common to all propagators momentum,
transforming the resulting integrals to integrals over Schwinger
parameters and then mapping these integrals to the required Feynman
integrals  by choosing auxiliary parameters from deformed propagators
lead to a functional equation.
  
  The following questions arise naturally:
  how to solve the functional equations 
  and how to use them for simplifying evaluation of Feynman integrals?
  We shall try to answer these questions in the next sections of this
  paper.

\subsection{Definitions and methods of solution}

A functional equation can be considered as an equation
involving independent variables, known functions, unknown functions
and some constants \cite{castillo2004functional}.
In a functional equation the unknown is a function.
Rather often, the functional equation connects a function with
its value for some other arguments.
There is a vast literature on  functional equations
\cite{aczel1989functional,
castillo2004functional,10.2307/43667209,
small2006functional,
rassias2000functional,efthimiou2011introduction,aczel1966lectures}.
Solution of a functional equation is a rather difficult problem.
However,  there is a number of the most frequently used methods for
its solution.
A systematic description of such methods is given in Ref. 
\cite{castillo2004functional}.
Many  methods described in  this book and also in Refs.
 \cite{aczel1989functional},  \cite{aczel1966lectures}
can be used for solving the functional equations for Feynman 
integrals. To our opinion, the most suitable methods are
\begin{itemize}
\item[1.]  Replacement of variables by given values 
\item[2.] Transforming one or several variables
\item[3.] Using a more general equation 
\item[4.] Treating some variables as constants
\item[5.] Iterative methods
\item[6.] Reduction by means of analytical techniques
(differentiation, integration etc.)
\item[6.] Mixed methods
\end{itemize}
All these methods to some extent can be used for solving
functional equations for Feynman integrals.
In the present paper, the methods 1, 3 and 5 will be exploited.

%%%%%%%%%%%%%%%%%%%%%%%%%%%%%%%%%%%%%%%%%%%%%%%%%%%%%%%%%%%%%%
%%%%%%%%%%%%%%%%%%%%%%%%%%%%%%%%%%%%%%%%%%%%%%%%%%%%%%%%%%%%%%

\subsection{Solution of the functional equation for the
propagator  integral}

%%%%%%%%%%%%%%%%%%%%%%%%%%%%%%%%%%%%%%%%%%%%%%%%%%%%%%%%%%%%%%
%%%%%%%%%%%%%%%%%%%%%%%%%%%%%%%%%%%%%%%%%%%%%%%%%%%%%%%%%%%%%%

%One of this methods will be used in our paper.
As an illustration of the first method from the above list,
we shall consider the solution of the functional equation
for the one-loop scalar propagator integral:
\begin{equation}
I_2^{(d)}(m_i^2,m_j^2;~s_{ij})=
\int \frac{d^d k_1}{i \pi^{{d}/{2}}}
\frac{1}{[(k_1-p_i)^2-m_i^2+i\eta]
         [(k_1-p_j)^2-m_j^2+i\eta]},
\end{equation}
where $i\eta$ is the small imaginary part which fixes the 
analytic continuation of the integral.
In Refs.  \cite{Tarasov:2008hw},\cite{Tarasov:2017},
the following relationship for this integral  was derived:
\begin{equation}
 I_2^{(d)}(m_i^2,m_j^2,s_{ij})=
 x_1I_2^{(d)}(m_j^2,m_0^2,s_{j0})
+x_2 I_2^{(d)}(m_i^2,m_0^2,s_{i0}),
\label{prop_fe}
\end{equation} 
where
\begin{eqnarray}
&&
x_1 = \frac{m_j^2-m_i^2+s_{ij}}{2s_{ij}}
 \pm \frac{\sqrt{4s_{ij}m_0^2-\lambda_{ij}}}{2 s_{ij}},
\nonumber \\
&&
x_2 = \frac{m_i^2-m_j^2+s_{ij}}{2s_{ij}} 
\mp \frac{\sqrt{4s_{ij}m_0^2-\lambda_{ij}}}{2 s_{ij}},
\label{solution_I2}
\end{eqnarray}
\begin{eqnarray}
&&s_{i0}=\frac{2s_{ij}(m_i^2+m_0^2)-\lambda_{ij}}{2s_{ij}}
\pm \frac{m_j^2-m_i^2-s_{ij}}{2s_{ij}} 
\sqrt{4s_{ij}m_0^2-\lambda_{ij}},
\nonumber \\
&&
\label{S13_S23}
\\
&&s_{j0}=\frac{2s_{ij}(m_j^2+m_0^2)-\lambda_{ij}}{2s_{ij}}
\pm \frac{m_j^2-m_i^2+s_{ij}}{2s_{ij}} 
\sqrt{4s_{ij}m_0^2-\lambda_{ij}}.
\nonumber
\end{eqnarray}
\begin{equation}
\lambda_{ij} = -s_{ij}^2-m_i^4-m_j^4+2 s_{ij}m_i^2
+2s_{ij}m_j^2+2m_i^2m_j^2.
\end{equation}
Equation (\ref{prop_fe})  can be interpreted  as a functional
equation for the integral $I_2^{(d)}(m_i^2,m_j^2;~s_{ij})$, 
which is considered  as a function of three continuous variables
$s_{ij}$, $m_i^2$, $m_j^2$.
To solve equation  (\ref{prop_fe}),  we will exploit a method,
which was used for the solution of Sincov's equation \cite{Sincov:1903a},
\cite{Sincov:1903b} :
\begin{equation}
f(x,y)=f(x,z)-f(y,z).
\label{fxy}
\end{equation}
Setting in this equation  $z=0$ and assuming that the 
function $f(x,z)$ is not singular at this point, we obtain the general 
solution
\begin{equation}
f(x,y)= g(y)-g(x),
\end{equation}
where
\begin{equation}
g(x)=f(x,0).
\end{equation}
Thus, using the fact that the left-hand side of  equation
 (\ref{fxy}) does not depend on $z$, we express the function
 $f(x,y)$  as a combination of its "boundary values".
 \\

 It is easy to see that the functional equation (\ref{prop_fe}) is rather 
 similar to  Sincov's equation (\ref{fxy}). 
Since at  $m_0^2=0$ the invariants  $s_{i0}$, $s_{j0}$ 
and the integral $I_2^{(d)}$  are not  singular,
one may set in  equation (\ref{prop_fe})   $m_0^2=0$. 
At $m_0^2=0$ equation (\ref{prop_fe}) becomes
\begin{eqnarray}
 I_2^{(d)}(m_i^2,m_j^2,s_{ij})=
 \overline{x}_1I_2^{(d)}(m_j^2,0,\overline{s}_{j0})
+\overline{x}_2 I_2^{(d)}(m_i^2,0,\overline{s}_{i0}),
\label{reshenie_fe_i2}
\end{eqnarray}
where
\begin{eqnarray}
&&
\overline{x}_1 = \frac{m_j^2-m_i^2+s_{ij}}{2s_{ij}}
 \pm \frac{\sqrt{-\lambda_{ij}}}{2 s_{ij}},
~~~~~~~~~
\overline{x}_2 = \frac{m_i^2-m_j^2+s_{ij}}{2s_{ij}} 
\mp \frac{\sqrt{-\lambda_{ij}}}{2 s_{ij}},
\label{x_solution_I2}
\end{eqnarray}
\begin{eqnarray}
&&\overline{s}_{i0}=\frac{2s_{ij}m_i^2-\lambda_{ij}}{2s_{ij}}
\pm \frac{m_j^2-m_i^2-s_{ij}}{2s_{ij}} 
\sqrt{-\lambda_{ij}},
\nonumber \\
&&\overline{s}_{j0}=\frac{2s_{ij}m_j^2-\lambda_{ij}}{2s_{ij}}
\pm \frac{m_j^2-m_i^2+s_{ij}}{2s_{ij}} 
\sqrt{-\lambda_{ij}}.
\label{sij_solution}
\end{eqnarray}
Therefore,  relation (\ref{reshenie_fe_i2})  
represents the integral depending on three variables
in terms of integrals depending on two variables. \\

%%%%%%%%%%%%%%%%%%%%%%%%%%%%%%%%%%%%%%%%%%%%%%%%%%%%%%%%%%%%%%%%%%
Expression (\ref{reshenie_fe_i2}) is a solution of  equation
(\ref{prop_fe}) for arbitrary value of the mass $m_0^2$.
Indeed, substituting (\ref{reshenie_fe_i2}) in both sides 
of equation (\ref{prop_fe}), simplifying arguments (\ref{sij_solution})
of integrals, after simple algebraic
transformations, we find that on the right-hand side the
integrals $I_2^{(d)}$ depending on $m_0^2$  are canceled.
The remaining two terms on the right-hand side are  
canceled  by the two terms from the left-hand side.  
\\

Notice that to reduce Feynman integrals to  simpler ones,
the question whether  expression (\ref{reshenie_fe_i2}) is 
a general solution of the functional equation (\ref{prop_fe})
or not is not relevant. For our purposes it is enough to have
a particular solution reducing complicated integral to a combination
of integrals with fewer variables.  Other sets of particular
solutions will lead to another representation of 
the complicated integral in terms of simpler ones.
These sets of integrals may be related, for example,
by analytic continuation or some transformation
analogous to the known transformations for hypergeometric
functions.\\

It should be noted that  $x_j$, $s_{j0}$ in equations
(\ref{solution_I2}), (\ref{S13_S23}) and  $\overline{x}_j$, 
$\overline{s}_{j0}$ in equations  (\ref{x_solution_I2}),(\ref{sij_solution}) 
do not depend on $i\eta$  that   can lead to  ambiguity 
in choosing the sign  of the square root.
Nevertheless, the functional equation will be valid for any choice 
of sign. The signs in $x_j$, $\overline{x}_j$   are to be properly correlated 
with the sings in $s_{j0}$, $\overline{s}_{j0}$. 
The possibility to choose different signs of the square root
means that there are two different representations of the 
integral in terms  of simpler integrals. The integrals on 
the right-hand side  of these two different  representations 
depend on different  sets of arguments. Excluding the initial 
integral from these  equations will give a functional equation 
for integrals with fewer arguments. As it was already
shown in Ref. \cite{Tarasov:2008hw}, this kind of functional
relations may be used for  the analytic continuation of integrals 
with  fewer variables.

We conclude this section with various remarks.
First, it is  interesting to note that the position of the 
threshold $s_{ij}=(m_i+m_j)^2$  for the integral on the left-hand 
side of equation (\ref{reshenie_fe_i2}) corresponds to
the positions of thresholds $s_{i0}=m_i^2$ and $s_{j0}=m_j^2$ 
for integrals on the right-hand side. \\

Second, we notice that the functional equations can be used 
not only for reducing complicated integrals to their "boundary 
integrals" but also   for analytic continuation of these "boundary
integrals". \\

Third,
to find an analytic expression for the simple integrals,
which cannot be simplified anymore by using functional
equations, one should use other computational methods.
In the next sections, we will use dimensional recurrence relations
for the triangle and box integrals at the final stage of calculation.

%%%%%%%%%%%%%%%%%%%%%%%%%%%%%%%%%%%%%%%%%%%%%%%%%%%%%%%%%%%%%%%%%%%%%%

\section{Functional reduction of the  integral $I_3^{(d)}$ }

%%%%%%%%%%%%%%%%%%%%%%%%%%%%%%%%%%%%%%%%%%%%%%%%%%%%%%%%%%%%%%%%%%%%%%

In the present paper, the functional equation and its solution
for  the one-loop triangle  integral 
with all internal  masses equal to zero will be considered. To solve 
the functional equation  for this integral,  we will use the functional 
equation for the  integral with massive lines:
\begin{equation}
I_3^{(d)}(m_1^2,m_2^2,m_3^2;s_{23},s_{13},s_{12})
=\frac{1}{i \pi^{d/2}} \int \frac{d^dk_1}{P_1P_2P_3},
\label{i3_definition}
\end{equation}
where 
\begin{equation}
P_i=(k_1-p_i)^2-m_i^2+i\eta.
\label{product3P}
\end{equation}

%%%%%%%%%%%%%%%%%%%%%%%%%%%%%%%%%%%%%%%%%%%%%%%%%%%%%%%%%%%%%%%%%%%%%%%%%%%%%%%%%%%%%%%%%%%%%
\subsection{Derivation of functional equation for the integral $I_3^{(d)}$  
            and its solution}
%%%%%%%%%%%%%%%%%%%%%%%%%%%%%%%%%%%%%%%%%%%%%%%%%%%%%%%%%%%%%%%%%%%%%%%%%%%%%%%%%%%%%%%%%%%%%

To derive the functional equation for the integral
(\ref{i3_definition}),  one can  exploit the 
algebraic relation for the  products of three propagators \cite{Tarasov:2017}:
\begin{equation}
\frac{1}{P_1 P_2 P_3}= \frac{x_{\scr 1}}{P_0 P_2 P_3 }
+\frac{x_{2}}{P_1 P_0 P_3}+\frac{x_{ 3}}{P_1 P_2  P_0}.
\label{3prop_relation}
\end{equation}
As it was shown in Ref. \cite{Tarasov:2017}, relation (\ref{3prop_relation}) is valid if
\begin{equation}
p_0 = x_1p_1+x_2p_2+x_3p_3,
\end{equation}
and the parameters $m_0^2$, $x_j$ obey the following system of equations:
\begin{eqnarray}
&&x_{1}+x_{2}+x_{ 3}=1,\\
&&x_{ 1}x_{ 2}s_{12}+x_{1}x_{\scr 3}s_{ 13}+x_{2}x_{ 3}s_{ 23}
-x_{1}m_1^2-x_{ 2}m_2^2-x_{3}m_3^2
 +m_{ 0}^2=0.
\end{eqnarray}
Solving  this system of equations for $x_{ 1}$,$x_{ 2}$, 
 we have
\begin{equation}
x_{\scr 1}=1-\Lambda_3-x_{ 3},~~~x_{ 2}=\Lambda_3,
\end{equation}
where  $\Lambda_3$ is the root of  the equation
\begin{equation}
A_3\Lambda_3^2 +B_3\Lambda_3 +C_3=0,
\end{equation}
with 
\vspace{-6mm}
\begin{eqnarray}
&&A_3=s_{\scr 12},\nonumber \\
&&B_3= x_{3}(s_{13}+s_{12}-s_{23})-m_1^2+m_2^2-s_{ 12},\nonumber \\
&&C_3= x_{3}^2s_{ 13} +(m_3^2-m_1^2-s_{13})x_{3}+m_1^2-m_{0}^2.
\end{eqnarray}
Integration of  relationship (\ref{3prop_relation}) with respect 
to the momentum $k_1$ gives a functional equation for the one-loop
integral $I_3^{(d)}$ with arbitrary masses:
\begin{eqnarray}
&&I_3^{(d)}(m_1^2,m_2^2,m_3^2;  s_{ 23},s_{ 13},s_{ 12})
=(1-\Lambda_3-x_{3})
I_3^{(d)}(m^2_{0},m_2^2,m_3^2;  s_{ 23},s_{ 30},s_{20})
\nonumber \\
&&\nonumber \\
&&~~~~~+\Lambda_3 
I_3^{(d)}(m_1^2,m_{0}^2,m_3^2;  {s}_{30},
{s}_{13}, s_{ 10})
+x_{3} I_3^{(d)}(m_1^2,m_2^2,m_{0}^2;  s_{20},{s}_{10},{s}_{12}),
\label{feI3massive}
\end{eqnarray}
where $s_{12}$, $s_{ 13}$, $s_{ 23}$  are independent scalar
invariants and   $s_{j0}$  are given by
\begin{eqnarray}
&&{s}_{10}=(p_1-p_0)^2=(m^2_1-m^2_2+s_{12})\Lambda_3+(m^2_1-m^2_3+s_{13})x_{3}
+m^2_0-m^2_1,
\nonumber \\
&&{s}_{20}=(p_2-p_0)^2=(m^2_1-s_{ 12}-m^2_2)\Lambda_3
+(m_1^2-m_3^2-s_{12}+s_{23})x_{3}+m^2_0-m^2_1+s_{12},
\nonumber \\
&&{s}_{30}=(p_3-p_0)^2=(m^2_1-s_{13}+s_{23}-m^2_2)\Lambda_3
+(m^2_1-m^2_3-s_{13})x_{3}+m^2_0-m^2_1+s_{13}.
\nonumber \\
&&
\label{i3_new_invars}
\end{eqnarray}

Setting $m_1=m_2=m_3=m_0=0$ in equations
(\ref{feI3massive}), (\ref{i3_new_invars}) and replacing
$s_{ij} \rightarrow q_{ij}$ in order to avoid confusion in the notation,
give the functional equation for the massless case
\begin{eqnarray}
&&I_3^{(d)}(0,0,0;  q_{ 23},q_{ 13},q_{ 12})
=(1-\lambda_3-z_{3})
I_3^{(d)}(0,0,0;  q_{ 23},q_{ 30}, {q}_{20})
\nonumber \\
&&\nonumber \\
&&~~~~~+\lambda_3 
I_3^{(d)}(0,0,0;  {q}_{30},
{q}_{13}, q_{ 10})
+z_{3} I_3^{(d)}(0,0,0;  q_{20},{q}_{10},{q}_{12}),
\label{feI3massless}
\end{eqnarray}
where
\begin{eqnarray}
&&{q}_{10}=q_{12}\lambda_3+q_{13}z_{3},
\nonumber \\
&&{q}_{20}= -q_{ 12}\lambda_3
+(q_{23}-q_{12})z_{3}+q_{12},
\nonumber \\
&&{q}_{30}=(q_{23}-q_{13})\lambda_3
-q_{13}z_{3}+q_{ 13},
\label{i3_new_invars_msls}
\end{eqnarray}
the parameter $z_3$ is arbitrary and  $\lambda_3$ 
is a solution of the quadratic equation:
\begin{equation}
q_{ 12} \lambda_3^2
+[z_{3}(q_{13}+q_{12}-q_{23})-q_{ 12}]\lambda_3+z_{3}(z_3-1)q_{ 13}=0.
\label{lam3_msls}
\end{equation}
The right- hand side of equation  (\ref{feI3massless}) 
depends on an arbitrary parameter $z_3$.
However, to exploit  this arbitrariness for obtaining a solution 
of the equation in terms of simpler integrals
by the method used for finding a solution of Sincov's equation
is not possible. One cannot  reduce the number of variables 
simultaneously in all functions $I_3^{(d)}$  by choosing $z_3$.
To find a solution of  equation (\ref{feI3massless}), 
one can use an approach described in \cite{castillo2004functional}.
Namely, we will find  a  solution for the integral with massless propagators 
from  a  more general functional equation which can be obtained from 
 formula (\ref{feI3massive}). As was  noted  in Ref. \cite{aczel1966lectures}, 
it can happen that
solving a more general equation   can be easier than 
solving a particular case of this equation.
Setting  $m_1^2=m_2^2=m_3^2=0$ in equation (\ref{feI3massive}) but
keeping $m_0$ different from zero, we find:
\begin{eqnarray}
&&I_3^{(d)}(0,0,0;  s_{ 23},s_{ 13},s_{ 12})
=(1-\overline{\Lambda}_3-x_{3})
I_3^{(d)}(m^2_{0},0,0; s_{ 23},s_{ 30},{s}_{20})
\nonumber \\
&&\nonumber \\
&&~~~~~+\overline{\Lambda}_3 
I_3^{(d)}(0,m_{0}^2,0; {s}_{30}, {s}_{13}, s_{ 10})
+x_{3} I_3^{(d)}(0,0,m_{0}^2; s_{20},{s}_{10},{s}_{12}),
\label{feI3mass_mm0}
\end{eqnarray}
where
\begin{eqnarray}
&&{s}_{10}= s_{12}\overline{\Lambda}_3+s_{13}x_{3}+m_0^2,
\nonumber \\
&&{s}_{20}=-s_{ 12}\overline{\Lambda}_3
+(s_{23}-s_{12})x_{3}+m^2_0+s_{12},
\nonumber \\
&&{s}_{30}=(s_{23}-s_{13})\overline{\Lambda}_3
-s_{13}x_{3}+m^2_0+s_{ 13},
\label{i3_new_invars_mm0}
\end{eqnarray}
and $\overline{\Lambda}_3$ is the solution of the equation:
\begin{equation}
s_{12} \overline{\Lambda}_3^2+[x_3(s_{13}+s_{12}-s_{23})-s_{12}]
\overline{\Lambda}_3 +x_3(x_3-1)s_{13} -m_0^2=0.
\end{equation}
We note that in the obtained equation there are more arbitrary parameters 
than in equation (\ref{feI3massless}) and also  new kinds of  integrals, namely, 
integrals with one massive internal line.  It is expected  that 
the solution  sought may be found by choosing 
arbitrary parameters $x_3$, $m_0$ .  Compared with  Sincov's equation, 
it is not so easy to  find  values of the arbitrary parameters leading 
to the reduction in the number of independent variables for all the integrals 
simultaneously. The reduction in the number of variables can take place if
by choosing parameters $x_3$, $m_0$, some of the variables
will become zero or equal to each other.
One can enumerate all   possible relations of this kind:
%%%%%%%%%%%%%%%%%%%%%%%
\begin{eqnarray}
&&s_{j0}=0,~~~s_{10}\pm s_{20}=0,~~~s_{10}\pm s_{30}=0,~~~s_{20}\pm s_{30}=0,
\nonumber \\
&&s_{j0}\pm m_0^2=0,~~~s_{j0}\pm s_{12}=0,
~~~s_{j0}\pm s_{13}=0 ,~~~s_{j0}\pm s_{23}=0.
\label{equs_for_fit_i3}
\end{eqnarray}
%%%%%%%%%%%%%%%%%%%%%%%
Probably there are some other conditions leading to the reduction
in the number of variables, but we will restrict ourselves only to the
conditions given in  equation (\ref{equs_for_fit_i3}).
Strictly speaking, at the expense of two parameters one can fulfill
two conditions from (\ref{equs_for_fit_i3}). Fulfilling  two conditions 
from the list (\ref{equs_for_fit_i3}) does not  ensure simultaneous reduction
in the number of variables for all three functions in  equation (\ref{feI3mass_mm0}).
However, we will try to find whether it is still  possible to choose
two arbitrary parameters and fulfill three conditions from the list
(\ref{equs_for_fit_i3}).
Out of  33 equations from the list (\ref{equs_for_fit_i3})  we created 5456 
different systems of equations with 3 equations in each system.
Solutions of these systems of equations and  analysis of these solutions 
were performed using  MAPLE.
It turns out that for some values of $m_0^2$ , $x_3$  three conditions
from the list (\ref{equs_for_fit_i3}) were fulfilled. 
In particular, one of such solution reads 
\begin{equation}
x_3= \overline{r}^{(3)}_{123},~~~~~~~~~~m_0^2=\mu_{123},
\label{solution_x3_m0}
\end{equation}
where
\begin{eqnarray}
&&
\mu_{ijk}=\frac{s_{ij}s_{ik}s_{jk}}
{s_{ij}^2+s_{ik}^2+s_{jk}^2
-2s_{ij}s_{ik}-2s_{ij}s_{jk}-2s_{ik}s_{jk}}
=\left. r_{ijk}\right|_{m_1^2=m_2^2=m_3^2=0}=\overline{r}_{ijk},
\nonumber 
\\
&&
\overline{r}^{(i)}_{jkl} = \left.\frac{\partial ~r_{jkl} }{\partial m_i^2} 
\right|_{m_j^2=m_k^2=m_l^2=0}.
\label{mu_ijk}
\end{eqnarray}
The definition of  $r_{ijk}$ is given in Appendix A.
For $x_3$, $m_0^2$ given in (\ref{solution_x3_m0}) the following conditions 
were fulfilled:
\begin{equation}
s_{10}=s_{20}=s_{30}= -m_{0}^2.
\end{equation}
Substituting these values of   $s_{j0}$ and 
$m_0$  into equation   (\ref{feI3mass_mm0}), we find 
\begin{eqnarray}
&&
I_3^{(d)}(0,0,0; ~s_{23},s_{13},s_{12})
\nonumber \\
&&~~~~~~~~~
= \overline{r}^{(1)}_{123} ~\xi_3^{(d)}(\overline{r}_{123}; s_{23}) 
+ \overline{r}^{(2)}_{123} ~\xi_3^{(d)}(\overline{r}_{123}; s_{13}) 
+ \overline{r}^{(3)}_{123} ~\xi_3^{(d)}(\overline{r}_{123}; s_{12}), 
\label{reshenie_ksi3}
\end{eqnarray}
where
\begin{eqnarray}
&&
\xi_3^{(d)}(\overline{r}_{ijk}; s_{ij})= 
I_3^{(d)}(0,0,\overline{r}_{ijk};
~-\overline{r}_{ijk},-\overline{r}_{ijk},s_{ij}),
\label{Qijk}
\end{eqnarray}
Explicit expressions for  $\overline{r}^{(i)}_{jkl}$  are given in  Appendix A.
Thus, the solution of the functional equation (\ref{feI3mass_mm0})  at
$m_1^2=m_2^2=m_3^2=0$  is a sum of three terms,
each of which is proportional to the same integral $\xi_3^{(d)}$ 
depending on two variables  different for each term.
We notice that $\mu_{ijk}$ is a kind of effective mass depending on  kinematic
invariants.
%%%%%%%%%%%%%%%%%%%%%%%%%%%%%%%%%%%%%%%%%%%%%%%%%%%%%%%%%%%%%%%%%%%%%

\subsection{Verification of the solution of the functional equation}

%%%%%%%%%%%%%%%%%%%%%%%%%%%%%%%%%%%%%%%%%%%%%%%%%%%%%%%%%%%%%%%%%%%%%
The obtained expression for the integral $I_3^{(d)}$ given in  equation 
(\ref{reshenie_ksi3}) is in fact the solution of the equation
(\ref{feI3massless}). 
In order to prove  this, we substitute the expression for $I_3^{(d)}$ from 
the equation  (\ref{reshenie_ksi3})  into  left- and right- hand sides 
of equation 
(\ref{feI3massless}). The  integrals $I_3^{(d)}$ on the right-hand side of 
(\ref{feI3massless}) depend on  different sets of variables, 
but it turns out that all these integrals are combinations of  
integrals $\xi_3^{(d)}$  depending on the same effective mass,
i.e. the following relations hold:
%%%%%%%%%%%%%%%%%%%%%%%%%%%%%%%%%%%%%%%%%%%%%%%
\begin{equation}
\mu_{123}=\mu_{023}=\mu_{103}=\mu_{120}.
\label{equal_masses_xi3}
\end{equation}
%%%%%%%%%%%%%%%%%%%%%%%%%%%%%%%%%%%%%%%%%%%%%%%%
Relations (\ref{equal_masses_xi3}) are easy to prove by replacing
$s_{ij}$ in   (\ref{mu_ijk}) by  $q_{ij}$ , taking into account   
(\ref{i3_new_invars_msls}) and using the relation in equation (\ref{lam3_msls}).
Due to  relations (\ref{equal_masses_xi3}) the number of 
$\xi_3^{(d)}$  functions in equation (\ref{feI3massless}) reduces from 12 to 6. 
On the right-hand side of this equation the contributions proportional to
$\xi_3^{(d)}(\mu_{123}; q_{10})$,
$\xi_3^{(d)}(\mu_{123}; q_{20})$,
$\xi_3^{(d)}(\mu_{123}; q_{30})$ drop out after  the algebraic
simplifications. The remaining contributions proportional to
$\xi_3^{(d)}(\mu_{123}; q_{23})$,
$\xi_3^{(d)}(\mu_{123}; q_{13})$,
$\xi_3^{(d)}(\mu_{123}; q_{12})$ 
cancel against the terms on the left-hand side of the equation.
Thus, having solved the functional equation, we expressed  the function
$I_3^{(d)}$
with three variables in terms of functions depending on two variables.
As we will see in the next subsection, the integral $\xi_3^{(d)}$  is easier  
to evaluate than the initial  integral. It should be noted that both the Feynman 
parameter representation for the integral $\xi_3^{(d)}$ 
 and the dimensional recurrence relation are simpler for this integral 
than for the initial integral $I_3^{(d)}$.

To conclude this section, we consider a particular case of equation
(\ref{reshenie_ksi3}), namely,  the case when one of the kinematic invariants, 
say $s_{12}$, is equal to zero. In this case, $I_3^{(d)}$ is  a linear 
combination of three  massless functions $\xi_3^{(d)}(0; s_{ij})$
\begin{equation}
I_3^{(d)}(0,0,0;  0,s_{13},s_{12})=
\frac{s_{13}}{s_{13}-s_{12}}~\xi_3^{(d)}(0; s_{13})
+\frac{s_{12}}{s_{12}-s_{13}}~\xi_3^{(d)}(0;  s_{12}).
\label{i3_part_case}
\end{equation}
The expression for $\xi_3^{(d)}(0;s_{ij})$ can be obtained from the recurrence
relation with respect to the spacetime dimension $d$, which
one can find, for example,  in Ref. \cite{Tarasov:1996br}:
\begin{equation}
(d-2)\xi_3^{(d+2)}(m^2; q^2)=
-2 \widetilde{m}^2 \xi_3^{(d)}(m^2; q^2)
-\xi_2^{(d)}(q^2),
\label{dim_rec_xi3}
\end{equation}
where
\begin{eqnarray}
&&\xi_2^{(d)}(q^2)=
\frac{1}{i \pi^{d/2}} \int \frac{d^dk_1}{[k_1^2+i\eta]
[(k_1-q)^2+i\eta]}=
 \frac{-\pi^{\frac32}~(-\widetilde{q}^{\: 2})^{\frac{d}{2}-2}}
{2^{d-3}  \Gamma\left(\frac{d-1}{2}\right)
 \sin \frac{\pi d }{2}},
\label{xi2_00q}
\end{eqnarray}
\begin{equation}
\widetilde{q}^{\: 2}=q^2+4i\eta,
\label{widetilde_qq}
\end{equation}
\begin{equation}
\widetilde{m}^{\: 2}=m^2-i\eta.
\label{widetilde_mm}
\end{equation}
We draw attention  to
the coefficient in front of the small imaginary part $i\eta$ in 
equation    (\ref{widetilde_qq}).  
The value of this coefficient is important for the
analytic continuation of  the results   in those cases  when  $I_2^{(d)}$ 
was used  in their derivation.
Expression (\ref{xi2_00q}) can be  considered as a limiting 
case of the propagator integral $I_2^{(d)}$ with equal internal masses.
Detailed derivation of   equation  (\ref{xi2_00q})  is given in Appendix B.
The importance of the coefficient in front of  $i\eta$  for 
the analytic continuation of the one-loop box integral  was noticed   
in Refs. \cite{Duplancic:2000sk}, \cite{Duplancic:2002dh}.

%%%%%%%%%%%%%%%%%%%%%%%%%%%%%%%%%%%%%%%%%%%%%
Setting $m^2=0$ in equation   (\ref{dim_rec_xi3})  and taking into account 
(\ref{xi2_00q}),  yield
\begin{equation}
\xi_3^{(d)}(0;q^2)=-\frac{1}{(d-4)}\xi_2^{(d-2)}(q^2)=
-\frac{2(d-3)}{q^2(d-4)}\xi_2^{(d)}(q^2).
\label{xi3mu3zero}
\end{equation}
Substituting  (\ref{xi3mu3zero}) in (\ref{i3_part_case}), we find:
\begin{equation}
I_3^{(d)}(0,0,0; 0,s_{13},s_{12})=
\frac{-2(d-3)}{(d-4)(s_{13}-s_{12})}\left[\xi_2^{(d)}(s_{13})-
\xi_2^{(d)}(s_{12})\right].
\label{i3_one_zero}
\end{equation}
Evaluating the Feynman parameter integral for the $I_3^{(d)}$
gives the same result. Expression (\ref{i3_one_zero}) will be used
in the next section for calculating the box type integral.

%%%%%%%%%%%%%%%%%%%%%%%%%%%%%%%%%%%%%%%%%%%%%%%%%%%%%%%%%%%%%
%%%%%%%%%%%%%%%%%%%%%%%%%%%%%%%%%%%%%%%%%%%%%%%%%%%%%%%%%%%%%

\subsection{Analytic evaluation of the integral $\xi_3^{(d)}$}

%%%%%%%%%%%%%%%%%%%%%%%%%%%%%%%%%%%%%%%%%%%%%%%%%%%%%%%%%%%%%
%%%%%%%%%%%%%%%%%%%%%%%%%%%%%%%%%%%%%%%%%%%%%%%%%%%%%%%%%%%%%
%%%%%%%%%%%%%%%%%%%%%%%%%%%%%%%%%%%%%%%%%%%%%%%%%%%%%%%%%%%%%
The analytic expression for the integral $\xi_3^{(d)}(m^2; q^2)$ 
can be derived by many different methods. For example, by direct evaluation
of the Feynman parameter integral
\begin{eqnarray}
&&\xi_3^{(d)}(m^2; q^2)
\nonumber \\
&&~~
=-\Gamma\left(3-\frac{d}{2}\right)
\int_0^1 \int_0^1
\frac{
z_{ 1} dz_{ 1}dz_{  2}}{\left[(q^2z_{ 2}^2-q^2z_{ 2}-m^2)
z_{ 1}^2 +m^2 - i\eta\right]^{3- \frac{d}{2}}},
\label{xi3_int_rep}
\end{eqnarray}
or by solving  dimensional  recurrence relation (\ref{dim_rec_xi3}) 
with respect to $d$. We prefer to use the latter method 
because  it is easier to keep the trace of the term $i \eta$ at all steps of
derivation.
To solve the dimensional recurrence  relation, we employ the  method 
described in Ref. \cite{Tarasov:2000sf}.
For  $|q^2| \leq |4m^2| $  the solution of the recurrence relation  
(\ref{dim_rec_xi3}) reads
\begin{equation}
\xi_3^{(d)}(m^2; q^2)=
-\frac{1}{2m^2}
\xi_2^{(d)}(q^2)
\Fh21\Ffw{1,\frac{d-2}{2}}{\frac{d-1}{2}}
+\frac{(-\widetilde{m}^2)^{d/2}}{\Gamma\left(\frac{d-2}{2}\right)}
C_3(q^2,d),
\label{ksi3_solutio_qq_small}
\end{equation}
where $C_3(q^2,d)$  is a periodic function  $C_3(q^2,d)=C_3(q^2,d+2)$. 
For this function one can obtain a differential equation with respect 
to $q^2$.
It can be derived from the differential equation  for $\xi_3^{(d)}(m^2, q^2)$.
In Ref. \cite{Tarasov:1996bz}, it was shown that the derivatives with respect
to kinematic  variables  for the Feynman integrals can be written in terms 
of integrals with shifted  space-time dimension $d$ and additional 
powers of propagators.  In our case,  the derivative reads:
\begin{eqnarray}
&&\frac{\partial}{\partial q^2} \xi_3^{(d)}(m^2; q^2)
\nonumber \\
&&~
=\frac{1}{ i\pi^{\frac{d+2}{2}}}
\int \frac{d^{d+2}k_1}
{((k_1-p_1)^2+i\eta)^2(k_1-p_2)^2+i\eta)^2 
((k_1-p_3)^2-m^2+i\eta)},
\label{qq_pro_ksi3}
\end{eqnarray}
where the values of  the kinematic invariants $s_{ij}= (p_i-p_j)^2$  must 
correspond to those  of the integral $I_3^{(d)}$ in (\ref{Qijk}).
With the help of the recurrence relations presented in Ref. \cite{Tarasov:1996br}
the integral on the right-hand side of (\ref{qq_pro_ksi3}) can be reduced to 
the set of basis integrals.  This reduction results in the 
first-order inhomogeneous differential equation:
\begin{eqnarray}
&&\frac{\partial}{\partial q^2} \xi_3^{(d)}(m^2; q^2)
=\frac{-(q^2+2m^2)}{q^2(q^2+4m^2)}
\xi_3^{(d)}(m^2; q^2)
\nonumber \\
&&~~~~~
-\frac{(d-3)}{\widetilde{q}^{\:2}(q^2+4m^2)} \xi_2^{(d)}(q^2)
  +\frac{d-2}{2 \widetilde{m}^2 q^2(q^2+4m^2)} \xi_1^{(d)}(m^2),
\label{ksi3_difequ}  
\end{eqnarray}
where
\begin{equation}
\xi_1^{(d)}(m^2)= \frac{1}{i\pi^{d/2}}\int \frac{d^dk_1}{k_1^2-m^2+i\eta}=
-\frac{\pi ( \widetilde{m}^2 )^{\frac{d}{2}-1} }
{\Gamma\left(\frac{d}{2}\right) \sin{\frac{\pi d}{2}}}.
\end{equation}
Substitution of  $\xi_3^{(d)}(m^2; q^2)$ from  (\ref{ksi3_solutio_qq_small}) 
into (\ref{ksi3_difequ}) yields
\begin{equation}
q^2\frac{\partial C_3(q^2,d)}{\partial q^2}
+\frac{(q^2+2m^2)}{q^2+4m^2} C_3(q^2,d)
+\frac{\Gamma\left(\frac{d}{2}\right)} { \left(-\widetilde{m}^2\right)^{d/2+1}
(q^2+4m^2)} \xi_1^{(d)}(m^2)=0.
\label{dif_equ_for_c3}
\end{equation}
Note that the term with  $\xi_1^{(d)}(m^2)$ in (\ref{dif_equ_for_c3})
is invariant with respect to the shift  $d \rightarrow d+2$, as it must be.
Equation  (\ref{dif_equ_for_c3}) can be solved by MAPLE. The solution
\begin{eqnarray}
&&
C_3(q^2,d)= \frac{ -\Gamma\left(\frac{d}{2}\right)\xi_1^{(d)}(m^2)}
{\sqrt{q^2(q^2+4m^2)} (-\widetilde{m}^2)^{d/2+1}}
 \ln\left(2m^2+q^2+ \sqrt{ (q^2+4m^2)q^2}\right)
\nonumber \\
&&~~~~~~~~~~~~~~~~~
 +\frac{K}{\sqrt{q^2(q^2+4m^2)}},
\label{reshenie_Cqq}
\end{eqnarray}
depends on a constant of integration  $K$, which may be fixed from
the comparison of  equation (\ref{ksi3_solutio_qq_small}) 
with the value of $\xi_3^{(d)}(m^2, q^2)$ taken at $q^2=0$. 
Substitution of   $q^2=0$ into the Feynman parameter integral 
(\ref{xi3_int_rep}) yields
\begin{equation}
\xi_3^{(d)}(m^2; 0)=-\Gamma\left(3-\frac{d}{2}\right)
\int_0^1 \frac{ z_1 dz_1 }
{ \left[(1-z_1^2)m^2-i\eta \right]^{3-\frac{d}{2}}}
=\frac{(d-2)}{4m^4}\xi_1^{(d)}(m^2).
\label{ksi3_qqzero}
\end{equation}
Since $\xi_3^{(d)}(m^2; 0)$  and the term with $_2F_1$ at $q^2=0$ 
are  finite,  the contribution proportional to the periodic 
function   $C_3(q^2,d)$  must also be finite. 
Taking the limit $q^2\rightarrow 0$ we find that the contribution 
coming from the term with $C_3(q^2,d)$  will be finite if
\begin{equation}
K=\frac{\Gamma\left(\frac{d}{2}\right)\xi_1^{(d)}(m^2)}
{(-\widetilde{m}^2)^{d/2+1}}\ln(2m^2).
\label{konstantaK}
\end{equation}
Substitution of (\ref{konstantaK}), (\ref{reshenie_Cqq}) in 
(\ref{ksi3_solutio_qq_small})  yields  
\begin{eqnarray}
&&
\xi_3^{(d)}(m^2; q^2)=
-\frac{1}{2m^2}\xi_2^{(d)}(q^2)
~\Fh21\Ffw{1,\frac{d-2}{2}}{\frac{d-1}{2}}
\nonumber \\
&& \nonumber \\
&&~~~~~~~~
+\frac{(d-2)\xi_1^{(d)}(m^2)}{2m^2\sqrt{q^2(q^2+4m^2)} }
 ~ \ln{\left(1+\frac{q^2+\sqrt{q^2(q^2+4m^2)}}{2m^2}\right)}.
\label{ksi3_arb_mom}
\end{eqnarray}
This expression is valid for $|\widetilde{q}^2/4/\widetilde{m}^2|<1$.
The results for $\xi_3^{(d)}(m^2; q^2)$ in  other kinematic  
regions can be related to (\ref{ksi3_arb_mom})
 by analytic continuations of the hypergeometric function $_2F_1$.

Relations (\ref{reshenie_ksi3}) and (\ref{ksi3_arb_mom})   have been 
checked for several values of kinematic variables by  numerical 
program~\cite{Borowka:2017idc} for evaluating  loop integrals.
We found  complete  agreement for both space-like and 
time-like values of the kinematic variables $s_{ij}$.

%%%%%%%%%%%%%%%%%%%%%%%%%%%%%%%%%%%%%%%%%%%%%%%%%%%%
\subsection{The $\varepsilon$ expansion of the integral $I_3^{(d)}$}
%%%%%%%%%%%%%%%%%%%%%%%%%%%%%%%%%%%%%%%%%%%%%%%%%%%%%

Analytic evaluation of the integral $I_3^{(d)}$ for various 
kinematic regions as well as its $\varepsilon=(4-d)/2$ expansion 
was considered in numerous papers 
\cite{Nickel:1978ds, Usyukina:1992jd,Lu:1992ny,Davydychev:1995mq, 
Davydychev:1997wa,Bern:1997sc,Davydychev:1999mq,
CabralRosetti:2002qv}.
To obtain $\varepsilon$ expansion of the integral $I_3^{(d)}$ 
depending on three variables, in our approach we need to know
$\varepsilon$ expansion of the integral $\xi_3^{(d)}$ 
depending only on two variables.
In order to get the first term in the $\varepsilon$ expansion of the integral 
$\xi_3^{(d)}$,   the hypergeometric function $_2F_1$ in (\ref{ksi3_arb_mom}) 
must be expanded  up to the  order $O(\varepsilon^2)$.
Using the HypExp package \cite{Huber:2005yg}, we find
\begin{eqnarray}
&&\Fh21\Fe{1,1-\varepsilon}{\frac32-\varepsilon}
= \frac{-Y}{(1-Y)(1+Y)}
\nonumber \\
&&~~~~\times  \left\{
2\ln (Y) - \varepsilon 
\left[ \ln^2 Y + 4\ln Y +4 {\rm Li}_2(1-Y)
\right]
\right\}+O(\varepsilon^2),
\label{expansionF21}
\end{eqnarray}
where
\begin{equation}
Y=\frac{1-y}{1+y},~~~~~~~~~~~~y=\sqrt{\frac{z}{z-1}}.
\end{equation}
Substituting (\ref{expansionF21}) into  (\ref{ksi3_arb_mom}),
we find
\begin{equation}
\xi_3^{(4-2\varepsilon)}(m^2; q^2)=
\frac{1}{2R}\left[\ln^2 Y + 2 \ln Y 
\ln \left(\frac{-\widetilde{q}^2}{\widetilde{m}^2}\right)
+4 {\rm Li}_2(1-Y) \right] + O(\varepsilon),
\label{xi3_eps_0}
\end{equation}
where
\begin{equation}
Y=1+\frac{\widetilde{q}^2+R}{2\widetilde{m}^2},
~~~~~~~~~~~~~~~
R=\sqrt{\widetilde{q}^2(\widetilde{q}^2+4\widetilde{m}^2)}.
\end{equation}
The leading term in $\varepsilon$ for the integral 
$I_3^{(d)}(0,0,0;s_{23},s_{13},s_{12})$ can now be obtained
by substituting (\ref{xi3_eps_0}) into equation (\ref{reshenie_ksi3}).
The resulting expression has been 
checked for several values of kinematic variables by  numerical 
program~\cite{Borowka:2017idc} for evaluating  loop integrals.
We found  complete  agreement for both space-like and 
time-like values of the kinematic variables $s_{ij}$.

%%%%%%%%%%%%%%%%%%%%%%%%%%%%%%%%%%%%%%%%%%%%%%%%%%%%%%%%%%%%%%%%%%%%%%%%%%
%%%%%%%%%%%%%%%%%%%%%%%%%%%%%%%%%%%%%%%%%%%%%%%%%%%%%%%%%%%%%%%%%%%%%%%%%%
%%%%%%%%%%%%%%%%%%%%%%%%%%%%%%%%%%%%%%%%%%%%%%%%%%%%%%%%%%%%%%%%%%%%%%%%%%
%%%%%%%%%%%%%%%%%%%%%%%%%%%%%%%%%%%%%%%%%%%%
\section{Functional reduction of the integral $I_4^{(d)}$ }
%%%%%%%%%%%%%%%%%%%%%%%%%%%%%%%%%%%%%%%%%%%
%with massless propagators}
In this section we will consider analytic evaluation of
the one-loop integral with massless internal lines associated with
the  Feynman  diagram with four external legs.
At $d=4$ the analytic result was presented in Refs.
\cite{Denner:1991qq,Usyukina:1993ch,Bern:1993kr}.
For particular values of kinematic variables,  analytic evaluation of this integral
for arbitrary  $d$ was considered, for
example,  in Refs. \cite{Bern:1993mq,Anastasiou:1999cx,
Duplancic:2000sk, Duplancic:2002dh}. 
Until now, in $d$-dimensions the analytic result  for this integral
for all external legs fully  off-shell   has been known in terms
of multiple hypergeometric series \cite{Davydychev:1990jt},
 \cite{Davydychev:1990cq}  evaluated by using the Mellin-Barnes technique.
In the present paper, such a result is derived  using a combination  of
functional equations and dimensional recurrence relations.

%%%%%%%%%%%%%%%%%%%%%%%%%%%
\subsection{Derivation of functional equation for the integral
$I_4^{(d)}$ and its solution}
%%%%%%%%%%%%%%%%%%%%%%%%%%%

In the present  paper, we will  be  concerned with the solution 
of the functional equation for the box integral  with massless 
internal propagators. 
However, to find a solution
for the integral with all massless internal propagators, we will use
the functional equation for the integral with massive internal lines.
For this reason,
we start our consideration with the functional equation for the 
box integral with all massive internal lines
\begin{equation}
I_4^{(d)}(m_1^2,m_2^2,m_3^2,m_4^2; 
s_{12},s_{23},s_{34},s_{14},s_{24},s_{13})
=\frac{1}{i \pi^{d/2}} \int \frac{d^dk_1}{P_1P_2P_3P_4},
\end{equation}
where
\begin{equation}
P_i=(k_1-p_i)^2-m_i^2+i\eta.
\label{product4P}
\end{equation}
The functional equation for this integral may  be obtained 
from the algebraic relation \cite{Tarasov:2017}:
\begin{equation}
\frac{1}{P_1 P_2 P_3 P_4}= \frac{x_{ 1}}{P_0 P_2 P_3 P_4}
+\frac{x_{2}}{P_1 P_0 P_3P_4}+\frac{x_{ 3}}{P_1 P_2  P_0P_4}
+\frac{x_{ 4}}{P_1 P_2  P_3P_0}.
\label{4prop_relation}
\end{equation}
Here the momentum $p_0$ is a combination of 
$p_1$, ..., $p_4$
\begin{equation}
p_0 = x_1p_1+x_2p_2+x_3p_3+x_4p_4,
\end{equation}
and  $m_0^2$, $x_j$ must satisfy the following conditions:
\begin{eqnarray}
&&x_{1}+x_{2}+x_{ 3}+x_4=1,\\
&&x_{ 1}x_{ 2}s_{12}+x_{1}x_{ 3}s_{ 13}+x_1x_4s_{14}
+x_{2}x_{ 3}s_{ 23}+x_2x_4s_{24}+x_3x_4s_{34}
\nonumber \\
&&
-x_{1}m_1^2-x_{ 2}m_2^2-x_{3}m_3^2-x_4 m_4^2
 +m_{ 0}^2=0.
\end{eqnarray}
The solution of this system for $x_{ 1}$,$x_{ 4}$ is
\begin{equation}
x_{ 1}=\Lambda_4,~~~x_{ 4}=1-x_2-x_3-\Lambda_4,
\end{equation}
where $\Lambda_4$ is a root of the quadratic equation,
\begin{equation}
A_4\Lambda_4^2 +B_4\Lambda_4 +C_4=0,
\label{Lambda_4}
\end{equation}
with
\vspace{-6mm}
\begin{eqnarray}
&&A_4=s_{\scr 14},\nonumber \\
&&B_4= (s_{24}-s_{12}+s_{14})x_2 +(s_{34}-s_{13}+s_{14})x_3
+m_1^2-m_4^2-s_{14},\nonumber \\
&&C_4= s_{24}x_2^2 +(s_{34}-s_{23}+s_{24})x_2x_3+(m_2^2-m_4^2-s_{24})x_2+s_{34}x_3^2
\nonumber \\
&&~~~~~~~~~~~~~~~~+(m_3^2-m_4^2-s_{34})x_3+m_4^2-m_0^2.
\label{coefs_equ_Lambda4}
\end{eqnarray}
Integrating relation (\ref{4prop_relation}) with respect to 
the momentum $k_1$, we obtain the functional equation for the 
one-loop box integral $I_4^{(d)}$ with massive internal
propagators
\begin{eqnarray}
&&I_4^{(d)}(m_1^2,m_2^2,m_3^2,m_4^2;
 s_{12},s_{23},s_{34},s_{14},s_{24},s_{13})
 \nonumber \\
&&=x_1I_4^{(d)}(m_0^2,m_2^2,m_3^2,m_4^2;
 s_{20},s_{23},s_{34},s_{40},s_{24},s_{30})
\nonumber \\
&&+x_2I_4^{(d)}(m_1^2,m_0^2,m_3^2,m_4^2;
 s_{10},s_{30},s_{34},s_{14},s_{40},s_{13})
\nonumber \\
&&+x_3
 I_4^{(d)}(m_1^2,m_2^2,m_0^2,m_4^2;
 s_{12},s_{20},s_{40},s_{14},s_{24},s_{10})
\nonumber \\
&&+x_4I_4^{(d)}(m_1^2,m_2^2,m_3^2,m_0^2;
 s_{12},s_{23},s_{30},s_{10},s_{20},s_{13}),
\label{feBox_massive}
\end{eqnarray} 
where
 $s_{12}$, $s_{23}$, $s_{34}$, $s_{14}$, $s_{24}$, $s_{13}$
are arbitrary scalar invariants, and $s_{i 0}$ are defined
by
\begin{eqnarray}
&&s_{10} = \Lambda_4 (m_4^2- s_{14}  - m_1^2 )
       + x_{3}  (m_4^2 - m_3^2  - s_{14} + s_{13}  )
       + x_{2}  (m_4^2 - m_2^2 - s_{14} + s_{12}  )
       \nonumber \\
&&~~~~~~~~~~~~~~~~~~~~~~~~~~~~~~~~~~~~~~~~~~~~~~~~~~       
       + s_{14} + m_0^2 - m_4^2,
\nonumber \\
&&s_{20} =
       \Lambda_4 (m_4^2 - m_1^2 - s_{24} + s_{12}  )
       + x_{3}  (m_4^2 - m_3^2  - s_{24} + s_{23}  )
       + x_{2}  (m_4^2 - m_2^2 - s_{24}  )
       \nonumber \\
&&~~~~~~~~~~~~~~~~~~~~~~~~~~~~~~~~~~~~~~~~~~~~~~~~~~       
              + s_{24} + m_0^2 - m_4^2,
\nonumber \\
&&s_{30} =
        \Lambda_4 (m_4^2 - m_1^2  - s_{34} + s_{13}  )
       + x_{3}  (m_4^2 - m_3^2- s_{34}  )
       + x_{2} (m_4^2 - m_2^2- s_{34} + s_{23}  )
       \nonumber \\
&&~~~~~~~~~~~~~~~~~~~~~~~~~~~~~~~~~~~~~~~~~~~~~~~~~~       
              + s_{34} + m_0^2 - m_4^2,
\nonumber \\
&&s_{40} =
        \Lambda_4 ( s_{14} + m_4^2 - m_1^2 )
       + x_{3}  ( s_{34} + m_4^2 - m_3^2 )
       + x_{2}  ( s_{24} + m_4^2 - m_2^2 )
       \nonumber \\
&&~~~~~~~~~~~~~~~~~~~~~~~~~~~~~~~~~~~~~~~~~~~~~~~~~~       
              + m_0^2 - m_4^2.
\label{si0_box}
\end{eqnarray}
Here   $m_0$, $x_{2}$, $x_{3}$  are arbitrary parameters.
Setting in
(\ref{feBox_massive}) and (\ref{si0_box}) all masses to zero and replacing 
$s_{ij} \rightarrow q_{ij}$, $x_i \rightarrow z_i$ give
\begin{eqnarray}
&&I_4^{(d)}(0,0,0,0;
 q_{12},q_{23},q_{34},q_{14},q_{24},q_{13})
 \nonumber \\
&&=z_1I_4^{(d)}(0,0,0,0;
 q_{20},q_{23},q_{34},q_{40},q_{24},q_{30})
\nonumber \\
&&+~z_2I_4^{(d)}(0,0,0,0;
 q_{10},q_{30},q_{34},q_{14},q_{40},q_{13})
\nonumber \\
&&+~z_3
 I_4^{(d)}(0,0,0,0;
 q_{12},q_{20},q_{40},q_{14},q_{24},q_{10})
\nonumber \\
&&+~z_4I_4^{(d)}(0,0,0,0;
 q_{12},q_{23},q_{30},q_{10},q_{20},q_{13}),
\label{feBox_msls}
\end{eqnarray} 
where $z_2$, $z_3$ are arbitrary parameters,
$q_{i 0}$ are defined as:
\begin{eqnarray}
&&q_{10} = -\lambda_4  q_{14}  
       + z_{3}  (q_{13}  - q_{14}  )
       + z_{2}  (q_{12}  - q_{14}  )
       + q_{14},
\nonumber \\
&&q_{20} =
       (q_{12}  - q_{24}  ) \lambda_4 
       + z_{3}  ( q_{23} - q_{24}  )
       - z_{2}   q_{24}  
       + q_{24},
\nonumber \\
&&q_{30} =
        ( q_{13}  - q_{34}  )\lambda_4 
       - z_{3}  q_{34}  
       + z_{2} (q_{23} - q_{34}  )
       + q_{34},
\nonumber \\
&&q_{40} =
        \lambda_4  q_{14}  
       + z_{3}  q_{34}  
       + z_{2}  q_{24},
\label{qi0_box}
\end{eqnarray}
and  $\lambda_4$ is the solution of the quadratic equation:
\begin{equation}
a_4 \lambda_4^2+b_4\lambda_4 + c_4=0, 
\end{equation}
with
\begin{eqnarray}
&&a_4=q_{14}, \nonumber \\
&&b_4=(q_{24}-q_{12}+q_{14})z_2 +(q_{34}-q_{13}+q_{14})z_3-q_{14},
\nonumber \\
&&c_4=q_{24}z_2^2 +(q_{34}-q_{23}+q_{24})z_2z_3 -q_{24}z_2 +q_{34}z_3^2-q_{34}z_3.
\end{eqnarray}
In order to find a solution of  equation (\ref{feBox_msls}), we
will exploit another functional equation, a more general one.
Such an equation will be obtained from  equation
(\ref{feBox_massive}), setting in it $m_1=m_2=m_3=m_4=0$
but retaining  $m_0$ different from zero. In this case,
\begin{eqnarray}
&&I_4^{(d)}(0,0,0,0;
 s_{12},s_{23},s_{34},s_{14},s_{24},s_{13})
 \nonumber \\
&&=x_1I_4^{(d)}(m_0^2,0,0,0;
 s_{20},s_{23},s_{34},s_{40},s_{24},s_{30})
\nonumber \\
&&+x_2I_4^{(d)}(0,m_0^2,0,0;
 s_{10},s_{30},s_{34},s_{14},s_{40},s_{13})
\nonumber \\
&&+x_3
 I_4^{(d)}(0,0,m_0^2,0;
 s_{12},s_{20},s_{40},s_{14},s_{24},s_{10})
\nonumber \\
&&+x_4I_4^{(d)}(0,0,0,m_0^2;
 s_{12},s_{23},s_{30},s_{10},s_{20},s_{13}),
\label{feBox_mm0}
\end{eqnarray} 
where
\begin{eqnarray}
&&s_{10} = s_{14} + m_0^2 -s_{14} \overline{\Lambda}_4 + (s_{12} - s_{14}) x_2 
+ (s_{13} - s_{14}) x_3,
\nonumber \\
&&
s_{20} = s_{24} + m_0^2 +(s_{12} - s_{24})  \overline{\Lambda}_4 - x_2 s_{24} 
+ (s_{23} - s_{24}) x_3,
\nonumber \\
&&
s_{30} = (s_{13} - s_{34})  \overline{\Lambda}_4  + s_{34} + m_0^2 + (s_{23} - s_{34}) x_2 
- s_{34} x_3,
\nonumber \\
&&
s_{40} =m_0^2 +s_{14}  \overline{\Lambda}_4  + s_{24} x_2 + s_{34} x_3,
\end{eqnarray}
and $\overline{\Lambda}_4$ is the solution of the quadratic equation
\begin{eqnarray}
&&s_{\scr 14} \overline{\Lambda}_4^2
+[(s_{24}-s_{12}+s_{14})x_2 +(s_{34}-s_{13}+s_{14})x_3-s_{14}]
\overline{\Lambda}_4
\nonumber \\
&&~~~~
+ s_{24}x_2^2 +(s_{34}-s_{23}+s_{24})x_2x_3-s_{24}x_2+s_{34}x_3^2
-s_{34}x_3-m_0^2=0.
\end{eqnarray}
In a manner similar to that for the integral $I_3^{(d)}$, 
we make a list of possible equations for invariants $s_{0j}$.
It turns out that by choosing $x_2$, $x_3$ and $m_0^2$, 
one may fulfill the following relations:
\begin{equation}
s_{10}=s_{20}=s_{30}=s_{40}=-m_0^2 =
\left. -r_{1234}\right|_{m_1=m_2=m_3=m_4=0}.
\label{equs_for_x2x3}
\end{equation}
Substituting  (\ref{equs_for_x2x3}) into equation (\ref{feBox_mm0}),
we find
\begin{eqnarray}
&&I_4^{(d)}(0,0,0,0; s_{12},s_{23},s_{34},s_{14},s_{24},s_{13}) 
\nonumber \\
&&~~~
 =\overline{r}^{(1)}_{1234} ~B_{234}^{(d)}(\mu_{4})
   +  \overline{r}^{(2)}_{1234}~B_{134}^{(d)}(\mu_{4})
   +  \overline{r}^{(3)}_{1234}~B_{124}^{(d)}(\mu_{4})
   +  \overline{r}^{(4)}_{1234}~B_{123}^{(d)}(\mu_{4}),
\label{I4vB4_via_r}
\end{eqnarray}
where
\begin{equation}
B_{ijk}^{(d)}(\mu_{4})=
  I_4^{(d)}(0,0,0,\mu_{4};  
  s_{ij},s_{jk},-\mu_{4},-\mu_{4},-\mu_{4},s_{ik}).
\end{equation}
\begin{eqnarray}
&&
\mu_4=
\left. r_{1234}\right|_{m_1=m_2=m_3=m_4=0}=\overline{r}_{1234}.
\nonumber \\
&& \nonumber \\
&&
\overline{r}^{(i)}_{jkls} = \left.\frac{\partial ~r_{jkls} }{\partial m_i^2} 
\right|_{m_j^2=m_k^2=m_l^2=m_s^2=0}.
\label{defi_pro_rjkl}
\end{eqnarray}
The explicit expressions for $\overline{r}_{jkls} $,
$\overline{r}^{(i)}_{jkls} $ are given in Appendix A.
Notice that the function  $B_{ijk}^{(d)}(\mu_{4})$ is totally
symmetric in $i, j, k$  and depends only on four
variables, namely  $\mu_4$, $s_{ij}$, $s_{jk}$ and $s_{ik}$.

At the next step we will try to reduce integrals $B_{ijk}^{(d)}$  
with four variables to a combination of integrals with
fewer variables. To achieve this goal, we will again use equation
(\ref{feBox_massive}). 
Setting in this equation
\begin{equation}
m_1^2=m_2^2=m_3^2=0,~~~~m_4^2=\mu_4,~~~~s_{34}=s_{14}=s_{24}=-\mu_4, 
\end{equation}
leads to the relation
\begin{eqnarray}
&&I_4^{(d)}(0,0,0,\mu_4;
 s_{ij},s_{jk},-\mu_4,-\mu_4,-\mu_4,s_{ik})
 \nonumber \\
&&=x_1I_4^{(d)}(m_0^2,0,0,\mu_4;
 s_{20},s_{jk},-\mu_4,s_{40},-\mu_4,s_{30})
\nonumber \\
&&+x_2I_4^{(d)}(0,m_0^2,0,\mu_4;
 s_{10},s_{30},-\mu_4,-\mu_4,s_{40},s_{ik})
\nonumber \\
&&+x_3
 I_4^{(d)}(0,0,m_0^2,\mu_4;
 s_{ij},s_{20},s_{40},-\mu_4,-\mu_4,s_{10})
\nonumber \\
&&+x_4I_4^{(d)}(0,0,0,m_0^2;
 s_{ij},s_{jk},s_{30},s_{10},s_{20},s_{ik}).
\label{feBox_step2}
\end{eqnarray} 
Here the parameters  $x_r$, $m_0^2$ are required to obey the following
conditions:
\begin{eqnarray}
&&x_{1}+x_{2}+x_{3}+x_4=1,
\nonumber
\\
&&x_{ 1}x_{ 2}s_{ij}+x_{1}x_{ 3}s_{ ik}
+x_{2}x_{ 3}s_{jk}-\mu_4 x_4 (2-x_4) +m_{ 0}^2=0.
\label{iksy_B4}
\end{eqnarray}
In order to find conditions on arbitrary parameters for
which the number of variables in all integrals on the right-hand
side of (\ref{feBox_step2}) is simultaneously diminishing, we will
follow the same strategy which was  employed for the functional reduction
of the integral $I_3^{(d)}$.
We have compiled  a list of equations similar to the list  (\ref{equs_for_fit_i3}). 
Out of equations from this list we made all possible systems of equations
consisting of 3 and 4 equations in each system.
Taking into account relations (\ref{iksy_B4}), all 
these systems were solved for $x_j$, $m_0^2$ by computer algebra system 
MAPLE. 
In particular, it was discovered that a reduction in the number 
of variables occurs at $x_4=0$ and   $s_{j0}$, $m_0^2$ given by
\begin{equation}
s_{10}=s_{20}=s_{30}=-m_0^2=-\mu_{ijk},~~~~~~ s_{40}=\mu_{ijk}-\mu_4.
\end{equation}
Substituting these values into  equation  (\ref{feBox_step2}),
we get
\begin{eqnarray}
&&\!\!
B_{ijk}^{(d)}(\mu_{4}) 
\nonumber \\
&&
=\overline{r}_{ijk}^{(i)}~ \xi_4^{(d)}(\mu_{ijk},\mu_{4}; s_{jk})
+\overline{r}_{ijk}^{(j)}~ \xi_4^{(d)}(\mu_{ijk},\mu_{4}; s_{ik})
+\overline{r}_{ijk}^{(k)}~ \xi_4^{(d)}(\mu_{ijk},\mu_{4}; s_{ij}),
\label{Bijk_via_xi4}
\end{eqnarray}
where
\begin{equation}
\xi_4^{(d)}(\mu_{ijk},\mu_{4}; s_{ij})
    =I_4^{(d)}(0, 0, \mu_{ijk}, \mu_{4}; 
    s_{ij}, -\mu_{ijk}, \mu_{ijk}-\mu_{4}, -\mu_{4}, -\mu_{4}, -\mu_{ijk}),
\label{xi4_via_I4}
\end{equation}
and  $\mu_{ijk}$ is defined in equation (\ref{mu_ijk}). The 
explicit expressions for  $\overline{r}^{(i)}_{jkl}$ are given 
in Appendix A. 
Thus, in equation (\ref{Bijk_via_xi4}) we achieved a reduction of the integral 
$B_{ijk}^{(d)}(\mu_4)$ with one massive internal line to a combination
of integrals depending on three kinematic variables.
Substituting  $B_{ijk}^{(d)}(\mu_4) $  from  equation
(\ref{Bijk_via_xi4})  into equation
(\ref{I4vB4_via_r}) yields
\begin{eqnarray}
&&I_4(0,0,0,0; s_{12},s_{23},s_{34},s_{14},s_{24},s_{13})
\nonumber \\
&&=
\overline{r}^{(1)}_{1234}
\left[ 
\overline{r}^{(2)}_{234} \xi_4^{(d)}(\mu_{234},\mu_4; s_{34})
+\overline{r}^{(3)}_{234} \xi_4^{(d)}(\mu_{234},\mu_4; s_{24})
+\overline{r}^{(4)}_{234} \xi_4^{(d)}(\mu_{234},\mu_4; s_{23})
\right]
\nonumber
\\
&&
+~\overline{r}^{(2)}_{1234}
\left[
\overline{r}^{(1)}_{134} \xi_4^{(d)}(\mu_{134},\mu_4; s_{34})
+\overline{r}^{(3)}_{134} \xi_4^{(d)}(\mu_{134},\mu_4; s_{14})
+\overline{r}^{(4)}_{134} \xi_4^{(d)}(\mu_{134},\mu_4; s_{13})
\right]
\nonumber \\
&&
+~\overline{r}^{(3)}_{1234}
\left[
\overline{r}^{(1)}_{124}\xi_4^{(d)}(\mu_{124},\mu_4; s_{24})
+\overline{r}^{(2)}_{124}\xi_4^{(d)}(\mu_{124},\mu_4; s_{14})
+\overline{r}^{(4)}_{124}\xi_4^{(d)}(\mu_{124},\mu_4; s_{12})\right]
\nonumber \\
&&
+~\overline{r}^{(4)}_{1234}\left[
 \overline{r}^{(1)}_{123}  \xi_4^{(d)}(\mu_{123},\mu_4; s_{23})
+\overline{r}^{(2)}_{123}  \xi_4^{(d)}(\mu_{123},\mu_4; s_{13})
+\overline{r}^{(3)}_{123}  \xi_4^{(d)}(\mu_{123},\mu_4; s_{12})
\right].
\nonumber \\
\label{Solu_for_I4msls}
\end{eqnarray}
We have not found  relationships reducing $\xi_4^{(d)}$  to  simpler 
integrals with fewer arguments.
Thus, using the two step functional reduction, we
expressed the integral depending on six variables
in terms of integrals $\xi_4^{(d)}$ depending only
on three variables. The analytic expression for 
the integral $\xi_4^{(d)}$ will be presented in  subsection
\ref{sec:evalxi4int}.
%%%%%%%%%%%%%%%%%%%%%%%%%%%%%%%%%%%%%%%%%%%%%%%%%%%%%%%%%%%%%%%%%%%%

\subsection{Verification of the solution of the functional equation}

%%%%%%%%%%%%%%%%%%%%%%%%%%%%%%%%%%%%%%%%%%%%%%%%%%%%%%%%%%%%%%%%%%%%
Now we will show that the obtained expression
(\ref{Solu_for_I4msls}) is the  solution of the functional
equation (\ref{feBox_msls}).
Substituting $I_4^{(d)}$  from (\ref{Solu_for_I4msls})
into the right- and left-hand sides of  relation
(\ref{feBox_msls}) we obtain 60 terms. 
The arguments of these functions are the ratios of rather big 
polynomials containing  various powers of  square roots of
some other polynomials.
However, after complicated simplifications of these arguments
they  became   rather compact and
it turns out that
the situation regarding  the integral $I_4^{(d)}$ is
similar to the case of the  integral $I_3^{(d)}$.  The effective masses
$\mu_{ijkr}$
for all the $\xi_4^{(d)}$ integrals  in  equation (\ref{feBox_msls})
are the same, i.e.
\begin{equation}
\mu_{1234}=\mu_{0234}=\mu_{1034}=\mu_{1204}=\mu_{1230}.
\end{equation}
On the right- hand side of the equation, 
after complicated algebraic simplifications of the coefficients in front of
the $\xi_4^{(d)}$ integrals, 40  terms with a
rather nontrivial dependence on the parameters  $z_2$, $z_3$, 
 cancel each other. 
The remaining 10 terms were exactly canceled  by 10 terms from  the left-hand side.
%%%%%%%%%%%%%%%%%%%%%%%%%%%%%%%%%%%%%%%%%%%%%%%%
%
\subsection{Reduction equations for $I_4^{(d)}$ with particular 
values of variables}
\label{sec:I4_partic_val}
%%%%%%%%%%%%%%%%%%%%%%%%%%%%%%%%%%%%%%%%%%%%%%%%
In practical applications the integral $I_4^{(d)}$ is needed for 
some kinematic variables $s_{ij}$ equal to zero \cite{Bern:1993kr}, 
\cite{Duplancic:2000sk,Glover:2001rd, Duplancic:2002dh,Kozlov:2016vqy, 
Chicherin:2017bxc}. 
For this reason it would be interesting to study possible simplifications
of  relation (\ref{Solu_for_I4msls}) for these particular values of
the kinematic  variables. In this section, we will use the shorthand
\begin{equation}
I_4^{(d)}(s_{12},s_{23},s_{34},s_{14},s_{24},s_{13})
\equiv I_4^{(d)}(0,0,0,0,; s_{12},s_{23},s_{34},s_{14},s_{24},s_{13}).
\end{equation}
\\

{\it a) The case  $ s_{12}=s_{23}=0$.  }
Substituting these values into  equation (\ref{Solu_for_I4msls}),
yields
\begin{eqnarray}
&&2(s_{13} s_{24}-s_{14} s_{24}+s_{14} s_{34}+s_{24}^2-s_{24} s_{34})
I_4^{(d)}(0,0,s_{34},s_{14},s_{24},s_{13})
\nonumber \\
&& =(s_{13} s_{24}-s_{14} s_{24}+2 s_{14} s_{34}-s_{24} s_{34})
\nonumber \\
&&\times \left[
 r_{134}^{(1)}\xi_4^{(d)}(\mu_{134},\mu_{2h}; s_{34})
+r_{134}^{(3)}\xi_4^{(d)}(\mu_{134},\mu_{2h}; s_{14})
+r_{134}^{(4)} \xi_4^{(d)}(\mu_{134},\mu_{2h}; s_{13})\right]
\nonumber \\
&&
+s_{13} s_{24} ~\xi_4^{(d)}(0,\mu_{2h}; s_{13})
-s_{14} s_{24} ~\xi_4^{(d)}(0,\mu_{2h}; s_{14})
\nonumber \\
&&
-s_{34} s_{24} ~\xi_4^{(d)}(0,\mu_{2h}; s_{34})
+2 s_{24}^2 ~\xi_4^{(d)}(0,\mu_{2h}; s_{24}),
\label{case_a}
\end{eqnarray}
where
\begin{equation}
\mu_{2h}=
\frac{-s_{13}s_{24}^2}{4(s_{13}s_{24}-s_{14}s_{24}
+s_{14}s_{34}+s_{24}^2-s_{24}s_{34})}.
\end{equation}
The considered case is the most complicated one.
 The number of terms in (\ref{case_a}) is less than in the general
 case but the remaining integrals are of the same complexity.

{\it b) The case  $s_{12}=s_{34}=0$.}
This case is simpler than the previous one.
\begin{eqnarray}
&&(s_{13}-s_{14}-s_{23}+s_{24})
I_4^{(d)}(0,s_{23},0,s_{14},s_{24},s_{13})
\nonumber \\
&&~~~~~~~~~~~~~ =
 s_{13} ~\xi_4^{(d)}(0,\mu_{2e}; s_{13})
-s_{14} ~\xi_4^{(d)}(0,\mu_{2e}; s_{14})
\nonumber \\
&& ~~~~~~~~~~~~~~
+s_{24} ~\xi_4^{(d)}(0,\mu_{2e}; s_{24})
-s_{23}~ \xi_4^{(d)}(0,\mu_{2e}; s_{23}),
\label{i4_s12_s34_zero}
\end{eqnarray}
where
\begin{equation}
\mu_{2e}=\frac{s_{14}s_{23}-s_{13}s_{24}}
{4(s_{13}-s_{14}-s_{23}+s_{24})}.
 \end{equation}
 As we will see in the next subsection each integral
 on the right-hand side of  (\ref{i4_s12_s34_zero}) is a combination 
 of Gauss hypergeometric functions.
\\ 

{\it c) The case $s_{12}=s_{23}=s_{34}=0$.} Substituting these
invariants $s_{ij}$
into  equation  (\ref{Solu_for_I4msls}), we find
\begin{eqnarray}
&&(s_{13}-s_{14}+s_{24})I_4^{(d)}(0,0,0,s_{14},s_{24},s_{13})
\nonumber \\
&&~~~~~~~= s_{13}\xi_4^{(d)}(0,\mu_{1m}; s_{13})
-s_{14} \xi_4^{(d)}(0,\mu_{1m}; s_{14})
+s_{24} \xi_4^{(d)}(0,\mu_{1m}; s_{24}),
\end{eqnarray}
where
\begin{equation}
\mu_{1m}=-\frac{s_{13}s_{24}}{4(s_{13}-s_{14}+s_{24})}.
\end{equation}
\\

{\it d) The case $s_{12}=s_{23}=s_{34}=s_{14}=0$.} Substitution
of these values into (\ref{Solu_for_I4msls}) yields
\begin{eqnarray}
&&I_4^{(d)}(0,0,0,0,s_{24},s_{13})
\nonumber \\
&&
= \frac{s_{13}}{(s_{13}+s_{24})} \xi_4^{(d)}(0,\mu_{0m}; s_{13})
+\frac{s_{24}}{(s_{13}+s_{24})} \xi_4^{(d)}(0,\mu_{0m}; s_{24}),
\label{box0m}
\end{eqnarray}
where
\begin{equation}
\mu_{0m}=
-\frac{s_{13}s_{24}}{4(s_{13}+s_{24})}.
\end{equation}
We see that in all the cases but a) the integrals $I_4^{(d)}$
are combinations of integrals $\xi_4^{(d)}(0,\mu_4,s_{ij})$ with
different arguments. The analytic expression for this integral
will be given in the next subsection.

%%%%%%%%%%%%%%%%%%%%%%%%%%%%%%%%%%%%%%%%%%%%%%%%%%%%%%%%%

\subsection{Analytic evaluation of the integral $\xi_4^{(d)}$}
\label{sec:evalxi4int}
%%%%%%%%%%%%%%%%%%%%%%%%%%%%%%%%%%%%%%%%%%%%%%%%%%%%%%%%%
An analytic result for the integral $\xi_4^{(d)}(\mu_{3}, \mu_{4}; s_{ij})$
can be derived by many different methods.
For example, it can be evaluated by direct integration 
of  the Feynman parameter integral
\begin{equation}
\xi_4^{(d)}(\mu_{3}, \mu_{4}; s_{ij})
=\Gamma\left(4-\frac{d}{2}\right) 
\int_0^1\int_0^1\int_0^1 dx_{1} dx_{2} dx_{3} 
x_{1}^2x_{2} h_4^{\frac{d}{2}-4},
\label{xi4_int_rep}
\end{equation}
where
\begin{equation}
h_4=x_{1}^2x_{2}^2x_{3}(x_{3}-1)s_{ij}
 -\mu_{3}x_{1}^2x_{2}^2
+(\mu_{3}-\mu_{4})x_{1}^2+\mu_{4}-i\eta.
\end{equation}
However, 
we prefer to exploit the recurrence
relation with respect to the space-time dimension $d$.
There are several reasons for using this method. 
First, it is easier to keep the trace of the 
small imaginary term $i\eta$; and  second,  the resulting 
expressions usually have a rather compact form in terms of rapidly
converging hypergeometric series.

The dimensional recurrence relation for the  integral
$\xi_4^{(d)}$  can be obtained from the results 
for the integral with the general kinematics  given in
Refs. \cite{Tarasov:1996br}, \cite{Fleischer:1999hq}
\begin{eqnarray}
&&(d-3)\xi_4^{(d+2)}(\mu_{ijk}, \mu_{4}; s_{ij})=
-2\mu_{4}
\xi_4^{(d)}(\mu_{ijk}, \mu_{4}; s_{ij})
-\xi_3^{(d)}(\mu_{ijk}; s_{ij}).
\label{xi4_dim_rec}
\end{eqnarray}

The solution of this equation can be obtained by
using the method described in Ref. \cite{Tarasov:2000sf}
and it reads
\begin{eqnarray}
&&\xi_4^{(d)}(\mu_3,{\mu}_4; s_{i j})=
\frac{1}{2}
\sum_{r=0}^{\infty}\frac{
\left(\frac{d-3}{2}\right)_r}
{(-\widetilde{\mu}_4)^{r+1} }
\xi_{3}^{(d+2r)}({\mu}_{3}; s_{i j})
+
\frac{(-\widetilde{\mu}_4)^{d/2}}
{\Gamma\left(\frac{d-3}{2}\right)} C_4(s_{i j},d).
\label{reshenie_ksi4}
\end{eqnarray}
Differentiating  relation (\ref{reshenie_ksi4}) with respect
to $s_{ij}$ and taking into account equation (\ref{ksi3_difequ}),
we obtain
\begin{eqnarray}
&&
 s_{ij}\frac{\partial \xi_4^{(d)}(\mu_3,{\mu}_4; s_{ij})}
{\partial s_{i j}}=
-\frac{(s_{i j}+2{\mu}_3)}
{(s_{i j}+4{\mu}_3)}
\xi_4^{(d)}(\mu_3,{\mu}_4; s_{i j})
+\frac{2(d-3)}{(s_{i j}+4{\mu}_3)(s_{i j}+4{\mu}_4)  }
\xi_2^{(d)}(s_{i j})
\nonumber 
\\
&& \nonumber \\
&&~~~~~~~~~~~~~~~~~~~~~+\frac{(d-2)}{4\mu_3\mu_4(s_{i j}+4{\mu}_3)}
\xi_1^{(d)}(\mu_3)
\Fh21\FMM{1,\frac{d-3}{2}}{\frac{d-2}{2}}
\nonumber 
\\
&& \nonumber \\
&&~~~~~~~~~~~~~~~~~~~~~
+\frac{(-\widetilde{\mu}_4)^{d/2}}
{\Gamma\left(\frac{d-3}{2}\right)}
\left [ s_{i j} \frac{\partial C_4(s_{i j},d)}
{\partial s_{ij} }
+  \frac{(s_{i j}+2{\mu}_3)}{(s_{i j}+4{\mu}_3)}
C_4(s_{i j},d)
\right].
\label{xi4_summa}
\end{eqnarray}
On the other hand, 
we can write this derivative of $\xi_4^{(d)}$ with respect to $s_{ij}$
in terms of the $d+2$ dimensional integral with additional powers
of propagators \cite{Tarasov:1996bz}
\begin{eqnarray}
&&
\frac{\partial }{\partial s_{ij}}
\xi_4^{(d)}(\mu_3,{\mu}_4; s_{i j})
=\frac{1}{i \pi^{\frac{d+2}{2}}}
\int \frac{d^{d+2}k_1}{P_1^2P_2^2P_3P_4},
\end{eqnarray}
where
\begin{eqnarray}
&&P_1 =(k_1-p_1)^2+i\eta, ~~~~P_2= (k_1-p_2)^2+i\eta, \nonumber \\
&&P_3=(k_1-p_3)^2-\mu_3+i\eta,~~~P_4=(k_1-p_4)^2-\mu_4+i\eta.
\label{P_in_proiz_xi4}
\end{eqnarray}
The kinematic invariants  $s_{ij}=(p_i-p_j)^2$ in  (\ref{P_in_proiz_xi4})
are to be the same as those for the integral (\ref{xi4_via_I4}).
After applying the recurrence relations 
\cite{Tarasov:1996br}, \cite{Fleischer:1999hq}
to reduce this integral to a set of basis integrals,
we obtain for $\xi_4^{(d)}$ the first-order differential equation 
\begin{eqnarray}
&&
s_{ij}(s_{ij}+4\mu_3)\frac{\partial}{\partial s_{ij}}
\xi_4^{(d)}(\mu_{3}, \mu_{4}; s_{ij})=
-(s_{ij}+2\mu_3)
\xi_4^{(d)}(\mu_{3}, \mu_{4}; s_{ij})
\nonumber
\\
&&
+\frac{2s_{ij}(d-3)}{(s_{ ij}+ 4\mu_4)
\widetilde{s}_{ij}}
\xi_2^{(d)}(s_{ij})
-\frac{2\mu_4(d-3)}{\widetilde{\mu}_4(s_{ ij}+4\mu_4)}
I_2^{(d)}(0, {\mu_4} ; -\mu_{ 4})
\nonumber \\
&&
+\frac{(d-3)}{2 \widetilde{\mu}_4}
I_2^{(d)}({\mu_3},{\mu_4};
{\mu}_3-{\mu}_4)
- \frac{(d-2)}{4\widetilde{\mu}_3\widetilde{\mu}_4 }
\xi_1^{(d)}(\mu_3).
\label{dif_ur_c4}
\end{eqnarray}
The integrals  $I_2^{(d)}$ in this formula can be simplified
by employing the functional equation derived in Ref.
\cite{Tarasov:2011zz}
\begin{eqnarray}
&&
I_2^{(d)}(m_1^2,m_2^2;s_{12})=
\frac{s_{12}-m_2^2+m_1^2}{2s_{12}}
I_2^{(d)}\left(m_1^2,m_1^2;\frac{(s_{12}-m_2^2+m_1^2)^2}{s_{12}}\right)
\nonumber \\
&&
~~~~~~~~~~~~~~~~~~~~~~+\frac{s_{12}-m_1^2+m_2^2}{2s_{12}}
I_2^{(d)}\left(m_2^2,m_2^2;\frac{(s_{12}-m_1^2+m_2^2)^2}{s_{12}}\right).
\label{funk_ura_i2}
\end{eqnarray}
Using this relation, we find
\begin{equation}
I_2^{(d)}({\mu}_3,{\mu}_4; \mu_3-\mu_4)
=I_2^{(d)}({\mu}_3,{\mu}_3; 4(\mu_3-\mu_4)),
\end{equation}
\begin{equation}
I_2^{(d)}(0,{\mu}_4; -\mu_4)
=I_2^{(d)}(0,0; -4\mu_4).
\end{equation}
The analytic result for the integral $I_2^{(d)}$ 
with equal masses is well known \cite{Bollini:1972bi}, 
\cite{Boos:1990rg}
\begin{equation}
I_2^{(d)}(m^2,m^2; q^2)
 =\frac{(d-2)}{2m^2} \xi_1^{(d)}(m^2)~ \Fh21\Ffp{1,2-\frac{d}{2}}{\frac32}.
\end{equation}

A combination of equations (\ref{xi4_summa}), (\ref{dif_ur_c4}) 
yields the first order differential equation for $C_4(s_{ij},d)$:
\begin{eqnarray}
&&
s_{ij}(s_{ij}+4\mu_3)\frac{\partial}{\partial s_{ij}}
C_4^{(d)}(s_{ij},d)
\nonumber \\
&&
-(s_{ij}+2\mu_3)C_4^{(d)}(s_{ij},d) - K_{4a}-\frac{K_{4b}}{s_{ij}+4\mu_4},
\label{dif_equ_for_C4}
\end{eqnarray}
where
\begin{eqnarray}
&&
K_{4a}=-\frac{\Gamma\left(\frac{d-1}{2}\right)}
{2\mu_4^2 (-\widetilde{\mu}_4)^{d/2}}~\xi_1^{(d)}(\mu_3)
\Fh21\FMM{1,\frac{d-1}{2}}{\frac{d}{2}}
+\frac{\Gamma\left(\frac{d-1}{2}\right)}
{(-\widetilde{\mu}_4)^{d/2+1}}~I_2^{(d)}(\mu_3,\mu_4;
\mu_3-\mu_4)
\nonumber \\
\nonumber \\
&&~~~~~~= \frac{-\pi^{1/2}\Gamma\left(\frac{d}{2}\right)}
{2(-\widetilde{\mu}_4)^{d/2+2} K
}
\xi_1^{(d)}(\mu_4),
\nonumber \\
&& \nonumber \\
&&K_{4b}
=\frac{4\Gamma\left(\frac{d-1}{2}\right)}
{(-\widetilde{\mu}_4)^{d/2}}~I_2^{(d)}(0,0;-4\mu_4)
,
~~~~~~~~~~~~~~~~~~
K=\sqrt{1-\frac{\widetilde{\mu}_3}{\widetilde{\mu}_4}}.
\end{eqnarray}
The solution of equation (\ref{dif_equ_for_C4}) is
\begin{equation}
C_4(s_{ij},d)=K_{4a} \varkappa_{a} + K_{4b}\varkappa_{b} 
+\frac{\varkappa_c}{R},
\end{equation}
where $\varkappa_c$ is a constant of integration,
\begin{equation}
\varkappa_a = -\frac{1}{R}
\ln \left( 2\mu_3+s_{ij}+{R} \right)
\end{equation}  
\begin{eqnarray}
&&\varkappa_b=-\frac{1}{R}
\int_0^{s_{ij}}
\frac{dx ~\sqrt{x(x+4\mu_3)}}{x(x+4\mu_3)(x+4\mu_4)}
\nonumber \\
&&~~~~=\frac{-1}{4\mu_4
\sqrt{\mu_{3}(s_{ij}+4\mu_3)}}
F_1\left(\frac12,1,\frac12,\frac32;-\frac{s_{ij}}{4\mu_4},
-\frac{s_{ij}}{4\mu_3}\right)
\nonumber \\
&&~~~~=\frac{1}{4\mu_4 (s_{ij}+4\mu_3)Z }
~\ln\left(\frac{1-Z}{1+Z}\right),
\end{eqnarray}
and
\begin{equation}
R=\sqrt{s_{ij}(s_{ij}+4\mu_3)},~~~~~~~~
Z=\left(\frac{s_{ij}(\mu_4-\mu_3)}{\mu_4(s_{ij}+4\mu_3)}
\right)^{1/2}.
\end{equation}
The constant of integration $\varkappa_c$ is easy to fix from the
finiteness condition for $C_4(s_{ij},d)$ as $s_{ij}\rightarrow 0$:
\begin{equation}
\varkappa_c = K_{4a} \ln 2\mu_3.
\end{equation}
Finally, combining  all  the contributions, we find
\begin{equation}
C_4(s_{ij},d)= \frac{\pi^{1/2}\Gamma\left(\frac{d}{2}\right)}
{2(-\widetilde{\mu}_4)^{d/2+2}} \xi_1(\mu_4)
\left[ \frac{ \ln \left( \frac{1-Z}{1+Z}\right)  }{(s_{ij}+4\mu_3)Z}
+ \frac{\ln \left(1+\frac{s_{ij}+R}{2\mu_3} \right)}
{K R} \right].
\label{C4_final}
\end{equation}

The analytic result for the first sum in  equation
(\ref{reshenie_ksi4}) can be obtained with the help of
equation (\ref{ksi3_arb_mom}), and it is
\begin{eqnarray}
&&\frac{1}{2}
\sum_{r=0}^{\infty}\frac{
\left(\frac{d-3}{2}\right)_r}
{(-\widetilde{\mu}_4)^{r+1} }
\xi_{3}^{(d+2r)}({\mu}_{3}; s_{i j})
=
-\frac{(d-2)}{4\widetilde{\mu}_3 \widetilde{\mu}_4 R}
\ln \left(1+\frac{s_{ij}+R}{2\mu_3} \right)
\xi_1^{(d)}(\widetilde{\mu}_3)
\Fh21\FMU{1,\frac{d-3}{2}}{\frac{d-2}{2}}
\nonumber \\
&& \nonumber \\
&&~~~~~~~
+\frac{1}{2 \widetilde{\mu}_3 \widetilde{\mu}_4 }
\left(\frac{\widetilde{\mu}_3}
{\widetilde{s}_{ij} + 4\widetilde{\mu}_3}\right)^{\!\!\frac12}
\xi_2^{(d)}(s_{ij})
F_1\left(\frac{d-3}{2},\frac12,1,\frac{d-1}{2};
\frac{-\widetilde{s}_{ij}}{4\widetilde{\mu}_3},
\frac{-\widetilde{s}_{ij}}{4\widetilde{\mu}_4}
\right).
\label{xi4_sum_ot_xi3}
\end{eqnarray}
Substituting (\ref{xi4_sum_ot_xi3}), (\ref{C4_final}) into
equation (\ref{reshenie_ksi4}) yields
\begin{eqnarray}
&&\xi_4^{(d)}(\mu_3,\mu_4; s_{ij})
\nonumber \\
&&
~~~=
 \frac{\pi^{\frac12}\Gamma\left(\frac{d}{2}\right)}
{2 {\mu}_4^{2}\Gamma\left(\frac{d-3}{2}\right)}  
\xi_1^{(d)}(\mu_4)
\left( \frac{ \ln \left( \frac{1-Z}{1+Z}\right)  }{(s_{ij}+4\mu_3)Z}
+ \frac{\ln \left(1+\frac{s_{ij}+R}{2\mu_3} \right)}
{K R} \right) 
\nonumber \\
&& \nonumber \\
&&~~~
-\frac{(d-2)}{4 {\mu}_3 {\mu}_4 R}
\xi_1^{(d)}({\mu}_3)
\Fh21\FMU{1,\frac{d-3}{2}}{\frac{d-2}{2}}
\ln \left(1+\frac{s_{ij}+R}{2\mu_3} \right)
\nonumber \\
&& \nonumber \\
&&~~~
+\frac{1}{2 {\mu}_3 {\mu}_4 }
\left(\frac{\widetilde{\mu}_3}
{\widetilde{s}_{ij} + 4\widetilde{\mu}_3}\right)^{\!\!\frac12}
\xi_2^{(d)}(s_{ij})
F_1\left(\frac{d-3}{2},\frac12,1,\frac{d-1}{2};
\frac{-\widetilde{s}_{ij}}{4\widetilde{\mu}_3},
\frac{-\widetilde{s}_{ij}}{4\widetilde{\mu}_4}
\right).
\label{xi4_final}
\end{eqnarray}
Performing an  analytic continuation of the 
hypergeometric function $_2F_1$ in (\ref{xi4_final}), we get 
\begin{eqnarray}
&&\xi_4^{(d)}(\mu_3,\mu_4; s_{ij})
=
 \frac{\pi^{\frac12}\Gamma\left(\frac{d}{2}\right)}
{2 {\mu}_4^{2}\Gamma\left(\frac{d-3}{2}\right)}  
\frac{\xi_1^{(d)}(\mu_4)  }{(s_{ij}+4\mu_3)Z}\ln \left( \frac{1-Z}{1+Z}\right) 
\nonumber \\
&& \nonumber \\
&&~~~
+\frac{(d-2)(d-4)}{4 {\mu}_3 {\mu}_4 R}
\xi_1^{(d)}({\mu}_3)
\Fh21\FCP{1,\frac{d-3}{2}}{\frac{3}{2}}
\ln \left(1+\frac{s_{ij}+R}{2\mu_3} \right)
\nonumber \\
&& \nonumber \\
&&~~~
+\frac{1}{2 {\mu}_3 {\mu}_4 }
\left(\frac{\widetilde{\mu}_3}
{\widetilde{s}_{ij} + 4\widetilde{\mu}_3}\right)^{\!\!\frac12}
\xi_2^{(d)}(s_{ij})
F_1\left(\frac{d-3}{2},\frac12,1,\frac{d-1}{2};
\frac{-\widetilde{s}_{ij}}{4\widetilde{\mu}_3},
\frac{-\widetilde{s}_{ij}}{4\widetilde{\mu}_4}
\right).
\label{Xi4_final}
\end{eqnarray}
This result is valid   in the region
\begin{equation}
\left|1-\frac{\widetilde{\mu_3}}{\widetilde{\mu_4}}
\right| <1,~~~~~
\left|\frac{\widetilde{s}_{ij}}{4\widetilde{\mu}_3}
\right|<1,~~~~
\left|\frac{\widetilde{s}_{ij}}{4\widetilde{\mu}_4}
\right|<1.
\end{equation}
From this expression we can obtain the value in any 
kinematic region by an analytic continuation of 
the ${_2F_1}$ Gauss hypergeometric function 
\cite{Bateman:100233} and the ${F_1}$ Appell hypergeometric function 
\cite{OlssonJMP5}, \cite{Bezrodnykh2017}.
\\

The terms with logarithms in (\ref{Xi4_final})
depend on the square roots $R$ and $Z$, which are independent of the small
imaginary addition $i\eta$. But this causes no problems
because the logarithms are multiplied by the factors $1/R$, $1/Z$,
and due to this fact it does not matter which sign to choose
for $R$ and $Z$. The sign should be taken the same
for the argument of the logarithm and for the factor
in front of the logarithm.
\\
Now let us consider the integral $\xi_4^{(d)}$ at $s_{12}=0$.
The value of this integral can be derived 
from  the Feynman parameter integral
\begin{eqnarray}
&&\xi_4^{(d)}(\mu_3, \mu_4; 0)=
-\frac{(d-2)}{8\widetilde{\mu}_3^2 \widetilde{\mu}_4}
\xi_1^{(d)}(\widetilde{\mu}_3)
\Fh21\FMU{1,\frac{d-3}{2}}{\frac{d-2}{2}}
\nonumber \\
&& \nonumber \\
&&~~~~~~~~~~~~~~~~~~~~~~~~~~~~
+\frac{\pi^{3/2}~ \widetilde{\mu}_4^{d/2-3}}
{4\mu_3 \Gamma\left(\frac{d-3}{2}\right) ~ \sin\frac{\pi d}{2}
 } \frac{K-1}{K}.
\end{eqnarray}
This result coincides with that obtained from (\ref{Xi4_final})
in the limit  $s_{ij} \rightarrow 0$.

At the end of this section, we present the analytic expression
for the integral $\xi_4^{(d)}(0, \mu_4,s_{ij})$, which emerged
in our consideration of particular cases of 
functional relations for the integral $I_4^{(d)}$.
The result for this integral can be obtained either as a solution
of the dimensional recurrence relation 
\begin{equation}
\xi_4^{(d+2)}(0, \mu_4; s_{ij})=-\frac{2 \widetilde{\mu}_4}{d-3}
\xi_4^{(d)}(0, \mu_4; s_{ij}) 
+ \frac{2}{s_{ij}(d-4)}\xi_2^{(d)}(s_{ij}),
\label{xi4mu3zero}
\end{equation}
which follows from equation (\ref{xi4_dim_rec}) taken
at $\mu_{ijk}=0$ and $\xi_3^{(d)}$ replaced by $\xi_2^{(d)}$,  
according to  equation (\ref{xi3mu3zero}),
or by performing an analytic continuation of the
relation  (\ref{Xi4_final}). In both cases the same result
was obtained
\begin{eqnarray}
&&\xi_4^{(d)}(0, \mu_4; s_{ij})
= \frac{(d-3)}{2 s_{ij} \mu_4} \xi_2^{(d)}(-4\mu_4) 
 \ln \left(1+\frac{\widetilde{s}_{ij}}{4\widetilde{\mu}_4} \right)
\nonumber \\
&& \nonumber \\
&&~~~~~~~~~~~~~~~~~~~~~~~~~
+\frac{(d-3)}{(d-4) s_{ij}\mu_4}
\xi_2^{(d)}(s_{ij}) 
\Fh21\Fsm{1,\frac{d-4}{2}}{\frac{d-2}{2}}.
\label{xi4_zero_mu3}
\end{eqnarray}
It is important to note  that for  evaluating   analytic 
results for all the three particular cases the only function, namely  
$\xi_4^{(d)}(0, \mu_4; s_{ij})$, is required.
This is one of the advantages of our functional reduction method.

%%%%%%%%%%%%%%%%%%%%%%%%%%%%%%%%%%%%%%%%%%%%%%%%

\subsection{The $\varepsilon$ expansion  of $I_4^{(d)}$ for particular 
values of variables}
%%%%%%%%%%%%%%%%%%%%%%%%%%%%%%%%%%%%%%%%%%%%%%%%

In this subsection, we will  present the leading term in the 
expansion of the integral $I_4^{(d)}$ for particular cases b), c) and d) 
considered  in  subsection (\ref{sec:I4_partic_val}).
We reserve derivation  of the $\varepsilon$ expansion of 
the integral $I_4^{(d)}$ for  general kinematics
for a future publication.

To obtain $\varepsilon=(4-d)/2$ expansion of $I_4^{(d)}$ for the cases
b), c) and d),  we need to know the first terms in the expansion only 
for the  integral $\xi_4^{(d)}(0, \mu_4,s_{ij})$. Plugging 
expansions for the $_2F_1$ function, obtained by using the 
HypExp package \cite{Huber:2005yg}, and for the integral $\xi_2^{(d)}$
into equation (\ref{xi4_zero_mu3}), we find
\begin{eqnarray}
&&
\xi_4^{(4-2\varepsilon)}(0, \mu_4; s_{ij})
\nonumber \\
&&~~=
-\frac{\Gamma(1+ \varepsilon) } {2\varepsilon^2 \mu_4 s_{ij}}
\left\{ 1-\varepsilon L_{ij}
-\varepsilon^2
\left[\frac{\pi^2}{3}-\frac12 L_{ij}^2
-{\rm Li}_2\left(1+\frac{\widetilde{s}_{ij}}{4\widetilde{\mu}_4}
      \right)\right] \right\}
      +O(\varepsilon).
\label{xi4dim4exp}
\end{eqnarray}
where
\begin{equation}
L_{ij}=\ln (-\widetilde{s}_{ij}).
\label{Lij}
\end{equation}
We compared numerical values derived from  this formula with the 
results obtained by the numerical program SecDec \cite{Borowka:2017idc}
and found complete agreement for both the space- and time-
like values of the scalar invariants  $s_{ij}$.

Using equation (\ref{xi4dim4exp}), it is easy to derive  from
equation (\ref{box0m}) the leading term in the $\varepsilon$
expansion for the integral $I_4^{(d)}$   for three particular cases
b), c) and d).

For the case b) we find:
\begin{eqnarray}
&&I_4^{(4-2\varepsilon )}(0,s_{23},0,s_{14},s_{24},s_{13})
=\frac{2}{(s_{14}s_{23}-s_{13}s_{24}) ~\varepsilon}
\left\{ L_{13}-L_{14}-L_{23}+L_{24} 
\right.
\nonumber \\
&& \nonumber \\
&&
+\varepsilon \left[ 
{\rm Li}_2\left(\!1+\!\frac{\widetilde{s}_{14}}{4\widetilde{\mu}_{2e}}\right)
\!-\!{\rm Li}_2\left(1\!+\!\frac{\widetilde{s}_{13}}{4\widetilde{\mu}_{2e}}\right)
\!-\!{\rm Li}_2\left(1\!+\!\frac{\widetilde{s}_{24}}{4\widetilde{\mu}_{2e}}\right)
\!+\!{\rm Li}_2\left(1\!+\!\frac{\widetilde{s}_{23}}{4\widetilde{\mu}_{2e}}\right)
\right.
\nonumber \\
&& \nonumber \\
&&\left.
\left.
~~~~~~~~~~~~~~~~~~~~~~
+\frac12\left( L_{14}^2- L_{13}^2- L_{24}^2+ L_{23}^2  \right)
 \right]\right\}.
\label{eps_case_b} 
\end{eqnarray}	  
Here and in formulae below we used  $L_{ij}$  defined in  equation (\ref{Lij}).

For the case c) the leading term in $\varepsilon$ is:
\begin{eqnarray}
&&I_4^{(4-2\varepsilon )}(0,0,0,s_{14},s_{24},s_{13})
=\frac{2 \Gamma(1+\varepsilon)}
{s_{13}s_{24}\varepsilon^2 }
\left\{1+\varepsilon (L_{14}-L_{13}-L_{24})
\right.
\nonumber \\
&&~~
 +\varepsilon^2 \left[
  {\rm Li}_2\left(1+ \frac{\widetilde{s}_{13}}{4\widetilde{\mu}_{1m}}\right)
 +{\rm Li}_2\left(1+ \frac{\widetilde{s}_{24}}{4\widetilde{\mu}_{1m}}\right)
 -{\rm Li}_2\left(1+ \frac{\widetilde{s}_{14}}{4\widetilde{\mu}_{1m}}\right)
\right.
\nonumber \\
&& \nonumber \\
&&\left. \left.
~~~~~~+\frac12 L_{13}^2-\frac12 L_{14}^2+\frac12 L_{24}^2
-2\zeta_2
 \right]
  \right\}.
\label{eps_case_c}
\end{eqnarray}

For the case c), when  all external legs are on-shell, we get
\begin{eqnarray}
&&I_4^{(4-2\varepsilon )}(0,0,0,0,s_{24},s_{13})
\nonumber \\
&&~~~~~~~
=\frac{4 \Gamma(1+\varepsilon)}
{s_{13}s_{24}\varepsilon^2 }\left\{
1-\frac{\varepsilon}{2}\left[L_{13}+L_{24}
\right]
+\frac{\varepsilon^2}{2}
\left[
L_{13} L_{24}  
- 5\zeta_2 \right]
\right\} + O(\varepsilon),
\label{I4onshell_eps_exp}
\end{eqnarray}
We compared the numerical results obtained 
from  (\ref{eps_case_b}),  (\ref{eps_case_c}),  (\ref{I4onshell_eps_exp})
and the results obtained by the numerical program SecDec  \cite{Borowka:2017}
and found perfect  agreement within the errors 
declared in this program for both  the time- and space- like 
invariants $s_{ij}$.

Next we turn to the derivation of the small $\varepsilon$ 
expansion for the integral $I_4^{(d)}$ taken at $d=6-2\varepsilon$.

From equations (\ref{xi4mu3zero}), (\ref{xi4dim4exp}) we obtain the first term 
in the expansion of the integral $\xi_4^{(6-2\varepsilon)}(0, \mu_4; s_{ij})$
for small  $\varepsilon=(6-d)/2$ 
\begin{equation}
\xi_4^{(6-2\varepsilon)}(0, \mu_4; s_{ij})
= \frac{1}{s_{ij}}\left[
{\rm Li}_2\left(1+\frac{\widetilde{s}_{ij}}{4\widetilde{\mu}_4}
            \right) 
            -\zeta_2 \right]
 +O(\varepsilon).
\label{xi4dim6exp}
\end{equation}

The integral $I_4^{(6-2\varepsilon )}$ for the case b) reads:
\begin{eqnarray}
&&I_4^{(6-2\varepsilon )}(0,s_{23},0,s_{14},s_{24},s_{13})
=\frac{1}{s_{13}-s_{14}-s_{23}+s_{24}}
\left[
{\rm Li}_2\left(1+\frac{\widetilde{s}_{13}}{4\widetilde{\mu}_{2e}}\right)
\right.
\nonumber \\
&& \nonumber \\
&&\left.
-{\rm Li}_2\left(1+\frac{\widetilde{s}_{14}}{4\widetilde{\mu}_{2e}}\right)
+{\rm Li}_2\left(1+\frac{\widetilde{s}_{24}}{4\widetilde{\mu}_{2e}}\right)
-{\rm Li}_2\left(1+\frac{\widetilde{s}_{23}}{4\widetilde{\mu}_{2e}}\right)
\right]
+O(\varepsilon).
\label{i4_6dim_case_b}
\end{eqnarray}
\vspace{0.3cm}
The integrals $I_4^{(6-2\varepsilon )}$ for the cases c) and d) read
\begin{eqnarray}
&&I_4^{(6-2\varepsilon )}(0,0,0,s_{14},s_{24},s_{13})
=\frac{-4 \widetilde{\mu}_{1m}}{s_{13}s_{24}}
\left[
{\rm Li}_2\left(1+\frac{\widetilde{s}_{13}}{4\widetilde{\mu}_{1m}}\right)
\right.
\nonumber \\
&&\left.~~~~~~~~~~~~~~~~~
-{\rm Li}_2\left(1+\frac{\widetilde{s}_{14}}{4\widetilde{\mu}_{1m}}\right)
+{\rm Li}_2\left(1+\frac{\widetilde{s}_{24}}{4\widetilde{\mu}_{1m}}\right)
-\zeta_2
\right]+O(\varepsilon),
\nonumber \\
&& \nonumber \\
&& \nonumber \\
&&
I_4^{(6-2\varepsilon )}(0,0,0,0,s_{24},s_{13})
=
\frac{-1}{2(s_{13}+s_{24})}
\left[ \ln^2 \left(\frac{\widetilde{s}_{24}}
{\widetilde{s}_{13}}\right) +6\zeta_2 \right]+O(\varepsilon).
\label{i4_6dim_case_g}
\end{eqnarray}
We notice that in all three cases the integrals have neither infrared 
nor ultraviolet divergences,  as it must be.
The result (\ref{i4_6dim_case_g}) agrees with the one presented in 
Ref. \cite{Bork:2013wga}.

We have check that the results for the $\varepsilon$ expansion
of the  $I_4^{(d)}$ integrals at $d=4-2\varepsilon$ and 
$d=6-2\varepsilon$ obey the following relations:
\begin{eqnarray}
&&
I_4^{(d+2 )}(0,0,0,0,s_{24},s_{13})
\nonumber \\
&&
=\frac{-2\widetilde{\mu}_{0m}}{(d-3)}~
I_4^{(d)}(0,0,0,0,s_{24},s_{13})
-\frac{8\widetilde{\mu}_{0m}}{(d-4)s_{13}s_{24}}
\left[\xi_2^{(d)}(s_{13})+ \xi_2^{(d)}(s_{24}) \right],
\end{eqnarray}

\begin{eqnarray}
&&
I_4^{(d+2 )}(0,0,0,s_{14},s_{24},s_{13})
=\frac{-2 \widetilde{\mu}_{1m}}{(d-3)}~
I_4^{(d)}(0,0,0,s_{14},s_{24},s_{13})
\nonumber \\
&&
~~~~~~~~~
-\frac{8\widetilde{\mu}_{1m} }{(d-4)s_{13}s_{24}}
\left[\xi_2^{(d)}(s_{13})- \xi_2^{(d)}(s_{14}) + \xi_2^{(d)}(s_{24})
\right],
\end{eqnarray}

\begin{eqnarray}
&&
I_4^{(d+2 )}(0,s_{23},0,s_{14},s_{24},s_{13})
=\frac{-2\widetilde{\mu}_{2e}}{(d-3)}
I_4^{(d)}(0,s_{23},0,s_{14},s_{24},s_{13})
\nonumber \\
&&
-\frac{8\widetilde{\mu}_{2e}}{(d-4)(s_{13}s_{24}-s_{14}s_{23})}
\left[\xi_2^{(d)}(s_{13})- \xi_2^{(d)}(s_{14})- \xi_2^{(d)}(s_{23})
+ \xi_2^{(d)}(s_{24}) \right],
\end{eqnarray}
where
\begin{eqnarray}
&&
\widetilde{\mu}_{0m}=\mu_{0m}-i\eta,
\nonumber \\
&&
\widetilde{\mu}_{1m}=\mu_{1m}-i\eta,
\nonumber \\
&&
\widetilde{\mu}_{2e}=\mu_{2e}
- i\eta.
\end{eqnarray}

%%%%%%%%%%%%%%%%%%%%%%%%%%%%%%%%%%%%%%%%%%%%%%%%%%%%%%%%%
%%%%%%%%%%%%%%%%%%%%%%%%%%%%%%%%%%%%%%%%%%%%%%%%%%%%%%%%%
\section{Conclusions and outlook}

%%%%%%%%%%%%%%%%%%%%%%%%%%%%%%%%%%%%%%%%%%%%%%%%%%%%%%%%%
%%%%%%%%%%%%%%%%%%%%%%%%%%%%%%%%%%%%%%%%%%%%%%%%%%%%%%%%%

 In this paper,  we have developed a systematic approach for calculating
 Feynman integrals with several kinematic variables and masses. 
 It is based on the iterative use of functional equations. 
 The functional equations  are used for reducing  Feynman integrals 
 to a combination of integrals with fewer variables.
 Integrals appearing after the last step of functional reduction
 were evaluated by using the method of dimensional recurrence relations
 developed in \cite{Tarasov:1996br, Tarasov:2000sf}.
 \\
   The approach  was applied for calculating one-loop triangle
   and box integrals with massless internal propagators.
   Our final reduction formulae for these integrals are given 
   in equations (\ref{reshenie_ksi3}), (\ref{Solu_for_I4msls}).   
   A striking  feature of  both relations is the fact that
   the integrals with massless
   internal propagators were expressed in terms of integrals
   with massive internal propagators. 
   Notice that at the second step of the functional reduction for the box 
   integral
   we performed  functional reduction already for the integral with a massive
   internal propagator.
   \\   
   
Integrals appearing  after the last iteration of the functional 
reduction were evaluated by using the dimensional recurrence relations.
A distinctive  feature of these recurrence relations is that
they are the first order inhomogeneous equations and the inhomogeneous
part has only one term. This significantly simplified their solution.  
   \\
   
There are many directions for future applications of the proposed
method. First of all, we are
going to apply our approach to the reduction of massless one-loop
 scalar integrals associated with diagrams with five and six external
 legs.
\\
Also, the method can be extended without problems to the one-loop 
integrals with  massive internal propagators.
\\

Another important direction for future  research  will  be the extension 
of the method of functional reduction to the evaluation of multiloop
integrals. One can easily write down functional equations
for multiloop integrals, but to elaborate a systematic
algorithm  one should solve a number of  problems. 
At the present time, we are  working on the solution of these problems.

\section{Acknowledgment}
I am grateful to Dmitri Kazakov for useful remarks concerning
the solution of functional equations for Feynman integrals.
This work was partly done during  the period  2012 -- 2016
and was supported by the German Research Foundation DFG 
within the  Collaborative Research Center
SFB 676 {\it Particles, Strings and the 
Early Universe: the Structure of Matter and Space-time}.

%%%%%%%%%%%%%%%%%%%%%%%%%%%%%%%%%%%%%%%%%%%%%%%%%%%%%%%%%%%%%%%%
\section{Appendix A}
%%%%%%%%%%%%%%%%%%%%%%%%%%%%%%%%%%%%%%%%%%%%%%%%%%%%%%%%%%%%%%%%

In this appendix, we  give the definition of the Gram determinants
and explicit formulae for polynomials occurring in equations
 (\ref{reshenie_ksi3}), (\ref{I4vB4_via_r}), (\ref{Solu_for_I4msls}).
 \begin{equation}
\Delta_n \equiv \Delta_n(\{p_1,m_1\},\ldots \{p_n,m_n\})=  \left|
\begin{array}{cccc}
Y_{11}  & Y_{12}  &\ldots & Y_{1n} \\
Y_{12}  & Y_{22}  &\ldots & Y_{2n} \\
\vdots  & \vdots  &\ddots & \vdots \\
Y_{1n}  & Y_{2n}  &\ldots & Y_{nn}
\end{array}
         \right|,~~~~~~~~~~~~~~~~~~~~~~
\label{deltan}
\end{equation}
\begin{equation}
Y_{ij}=m_i^2+m_j^2-s_{ij},
\end{equation}
\begin{eqnarray}
G_{n-1} \equiv G_{n-1}(p_1,\ldots ,p_n)= 
-2\left|
\begin{array}{cccc}
  \! S_{11}  & S_{12}  &\ldots & S_{1~ n-1}  \\
  \! S_{21}  & S_{22} &\ldots & S_{2~ n-1} \\
  \vdots  & \vdots  &\ddots & \vdots \\
  \! S_{n-1~ 1}  & S_{n-1 ~2} &\ldots & S_{n-1~ n-1}
\end{array}
\right|, ~~~
\end{eqnarray}
\begin{equation}
~S_{ij}=s_{i n}+s_{jn}-s_{ij},
\label{Gn}
\end{equation}
where
$s_{ij}^2=(p_i-p_j)^2$, 
$p_i$ are combinations of external momenta flowing through $i$-th
lines, respectively, and $m_i$ is the mass of the $i$-th line.
Where no confusion can arise, we simply refer to the above
functions as $\Delta_n$, $G_{n-1}$. 
We will also use  an indexed notation for $\Delta_n$ and $G_{n-1}$
%&&&&&&&&&&&&&&&&&&&&&&&&&&&&&&&&&&&&&&&&&&&&&&&&&&&&&&&&&&&&
\begin{eqnarray}
&&\lambda_{i_1 i_2 \ldots i_n } = 
 \Delta_n(\{p_{i_1},m_{i_1}\},\{p_{i_2},m_{i_2} \},
  \ldots ,\{p_{i_n},m_{i_n}\}),                         \nonumber \\
&& \nonumber \\
&&g_{i_1 i_2 \ldots i_n }=G_{n-1}(p_{i_1},p_{i_2},\ldots ,p_{i_n}).
\label{lage}
\end{eqnarray}
%&&&&&&&&&&&&&&&&&&&&&&&&&&&&&&&&&&&&&&&&&&&&&&&&&&&&&&&&&&&&
%Rather frequently the 
Our results  depend on the ratios of 
$\lambda_{i_1 i_2 \ldots i_n}$  and $g_{i_1 i_2 \ldots i_n}$  and,
therefore,  it is  convenient to introduce the notation
\begin{equation}
r_{ij\ldots k}=-\frac{\lambda_{ij\ldots k}}{g_{ij\ldots k}}.
\label{r_definition}
\end{equation}

The imaginary part of  $r$ is rather simple. Using
\begin{equation}
\sum_{j=1}^n \partial_j \lambda_{i_1{\ldots} i_n}
= -g_{i_1{\ldots} i_n}= -G_{n-1},
\label{Gnm1}
\end{equation}
one shows that to all orders in $\eta$
\begin{equation}
\lambda_{i_1 i_2 \ldots  i_n}(\{m_r^2-i\eta \})=
\lambda_{i_1 i_2 \ldots  i_n}(\{m_r^2\})+i g_{i_1 i_2 \ldots i_n} 
~ \eta,
\end{equation}
and, therefore, the causal $\eta$ prescription for $r$ is
(with the same $\eta$ for all masses)
\begin{equation}
\left. r_{ij\ldots k}\right|_{m_j^2-i\eta}=
\left. r_{ij\ldots k}\right|_{m_j^2}-i\eta.
\label{repsilon}
\end{equation}

For the reader's convenience we present below  explicit expressions
for the ratios of Gram determinants and their derivatives occurring  
in the reduction formulae for the integrals
$I_3^{(d)}$ and $I_4^{(d)}$:
\begin{eqnarray}
&&\overline{r}^{(i)}_{ijk}=\frac{2s_{jk}(s_{jk}-s_{ij}-s_{ik})}{g_{ijk}},
\nonumber \\
&&\overline{r}^{(j)}_{ijk}=\frac{2s_{ik}(s_{ik}-s_{ij}-s_{jk})}{g_{ijk}},
\nonumber \\
&&\overline{r}^{(k)}_{ijk}=\frac{2s_{ij}(s_{ij}-s_{ik}-s_{jk})}{g_{ijk}},
\nonumber \\
&& \widetilde{\overline{r}}_{njk}= - \frac{\overline{\lambda}_{njk}}
                           {g_{njk}} =		\overline{r}_{njk}-i \eta,
\end{eqnarray}
\begin{eqnarray}
&&
\overline{r}^{(i)}_{ijkn}=\frac{1}{g_{ijkn}}[
2 s_{ij} s_{jk} s_{kn}+2 s_{ij} s_{jn} s_{kn}-2 s_{ij} s_{kn}^2
+2 s_{ik} s_{jk} s_{jn}-2 s_{ik} s_{jn}^2
\nonumber \\
&&~~~~~~~~~~~~~~
+2 s_{ik} s_{jn} s_{kn}
-2 s_{jk}^2 s_{in}-4 s_{jk} s_{jn} s_{kn}+2 s_{jk} s_{jn}
 s_{in}+2 s_{jk} s_{kn} s_{in}],
\nonumber \\
&&
\overline{r}^{(j)}_{ijkn}=\frac{1}{g_{ijkn}}[
2 s_{ij} s_{ik} s_{kn}-2 s_{ij} s_{kn}^2+2 s_{ij} s_{kn} s_{in}
-2 s_{ik}^2 s_{jn}+2 s_{ik} s_{jk} s_{in}
\nonumber \\
&&~~~~~~~~~~~~~
+2 s_{ik} s_{jn} s_{kn}+2 s_{ik} s_{jn} s_{in}-4 s_{ik} s_{kn} s_{in}
+2 s_{jk} s_{kn} s_{in}-2 s_{jk} s_{in}^2],
\nonumber \\
&&
\overline{r}^{(k)}_{ijkn}=\frac{1}{g_{ijkn}}[2 s_{ij} s_{ik} s_{jn}
-2 s_{ij}^2 s_{kn}+2 s_{ij} s_{jk} s_{in}
+2 s_{ij} s_{jn} s_{kn}-4 s_{ij} s_{jn} s_{in}
\nonumber \\
&&~~~~~~~~~~~~~
+2 s_{ij} s_{kn} s_{in}-2 s_{ik} s_{jn}^2+2 s_{ik} s_{jn} s_{in}
+2 s_{jk} s_{jn} s_{in}-2 s_{jk} s_{in}^2],
\nonumber \\
&&
\overline{r}^{(n)}_{ijkn}=\frac{1}{g_{ijkn}}[2 s_{ij} s_{ik} s_{jn}
-2 s_{ij}^2 s_{kn}-4 s_{ij} s_{ik} s_{jk}
+2 s_{ij} s_{ik} s_{kn}+2 s_{ij} s_{jk} s_{kn}
\nonumber \\
&&~~~~~~~~~~~~~
+2 s_{ij} s_{jk} s_{in}-2 s_{ik}^2 s_{jn}+2 s_{ik} s_{jk} s_{jn}
+2 s_{ik} s_{jk} s_{in}-2 s_{jk}^2 s_{in}],
\nonumber \\
&& \widetilde{\overline{r}}_{ljkn}= - \frac{\overline{\lambda}_{ljkn}}
                                        {g_{ljkn}} =
					\overline{r}_{ljkn}-i\eta,
\end{eqnarray}

where
\begin{eqnarray}
&&
g_{njk}=2(s_{nj}^2+s_{nk}^2+s_{jk}^2-2s_{nj}s_{nk}-2s_{nj}s_{jk}-2s_{nk}s_{jk}),
\nonumber \\
&&
\overline{\lambda}_{njk}=-2s_{nj}s_{jk}s_{nk}+i\eta g_{njk},
\end{eqnarray}
\begin{eqnarray}
&&g_{ljkn}=
4 s_{lj}^2 s_{kn}+4 s_{lj} s_{lk} s_{jk}-4 s_{lj} s_{lk} s_{jn}
-4 s_{lj} s_{lk} s_{kn}
-4 s_{lj} s_{ln} s_{jk}
\nonumber \\
&&~~~~~
+4 s_{lj} s_{ln} s_{jn}
-4 s_{lj} s_{ln} s_{kn}-4 s_{lj} s_{jk} s_{kn}-4 s_{lj} s_{jn} s_{kn}
+4 s_{lj} s_{kn}^2
\nonumber \\
&&~~~~~
+4 s_{lk}^2 s_{jn}-4 s_{lk} 
s_{ln} s_{jk}-4 s_{lk} s_{ln} s_{jn}+4 s_{lk} s_{ln} s_{kn}-4 
s_{lk} s_{jk} s_{jn}
\nonumber \\
&&~~~~~
+4 s_{lk} s_{jn}^2-4 s_{lk} s_{jn} s_{kn}+
4 s_{ln}^2 s_{jk}+4 s_{ln} s_{jk}^2-4 s_{ln} s_{jk} s_{jn}
\nonumber \\
&&~~~~~
-4 
s_{ln} s_{jk} s_{kn}+4 s_{jk} s_{jn} s_{kn},
\end{eqnarray}
\begin{eqnarray}
&&
\overline{\lambda}_{ljkn}
=
-2 s_{lj} s_{lk} s_{jn} s_{kn}
-2 s_{lj} s_{ln} s_{jk} s_{kn}
-2 s_{lk} s_{ln} s_{jk} s_{jn}
\nonumber \\
&&~~~~~~~~~~~~~
+s_{lj}^2 s_{kn}^2
+s_{lk}^2 s_{jn}^2
+s_{ln}^2 s_{jk}^2
+i\eta g_{ljkn}.
\end{eqnarray}

\section{Appendix B}
In this appendix, we  describe derivation of the analytic dependence
of the one-loop massless propagator integral 
(\ref{xi2_00q}) on the small imaginary
part $i\eta$ added to the propagators. 
Expression (\ref{xi2_00q}) can be obtained as a leading term 
of the analytic result for the integral
%&&&&&&&&&&&&&&&&&&&&&&&&&&&&&&&&&&&&&&&&&&&&&&&&&&&&&&&&&&&&
\begin{eqnarray}
I_2^{(d)}(m^2,m^2; q^2) =
\int \frac{d^d k_1}{[i \pi^{{d}/{2}}]} 
\frac{1}{((k_1-q)^2-m^2+i\eta)(k_1^2-m^2+i\eta)},
\label{i2_mm}
\end{eqnarray}
%&&&&&&&&&&&&&&&&&&&&&&&&&&&&&&&&&&&&&&&&&&&&&&&&&&&&&&&&&&&&
taken at $m^2=0$ and $\eta \rightarrow 0$.
The integral (\ref{i2_mm})
can be obtained as a solution of the dimensional recurrence
relation
\begin{eqnarray}
2(d-1)I_2^{(d+2)}(m^2,m^2; q^2 ) 
-(q^2-4\widetilde{m}^2)
I_2^{(d)}(m^2,m^2; q^2 )
+2\xi_1^{(d)}(m^2) =  0.
\end{eqnarray}
At  $|q^2|>4|\widetilde{m}^2|$  the solution of this equation reads 
\cite{Tarasov:1996br}
\begin{eqnarray}
&&I_2^{(d)}(m^2,m^2; q^2 )=\frac{-\pi^{3/2}}{  2^{d-3}
\Gamma\left(\frac{d-1}{2}\right)q^4 \sin \frac{\pi d}{2} }
\left(\frac{q^2}{q^2-4\widetilde{m}^2}\right)^{\frac32}
\left(-q^2+4\widetilde{m}^2\right)^{\frac{d}{2}}
\nonumber \\
&& \nonumber \\
&&~~~~~~~~~~~~~~~~~~~~~~~~~~~~~~~~~~~~~~
-\frac{2\pi~ \widetilde{m }^{ d-2}}
{q^2 \Gamma\left(\frac{d}{2}\right) \sin \frac{\pi d}{2}}
\Fh21\FmS{1,\frac12}{\frac{d}{2}}.
\label{i2_eqmass}
\end{eqnarray}
This expression agrees with the one presented in   Ref. \cite{Boos:1990rg}.
For the massless case in expression (\ref{i2_eqmass}) we must set 
$m^2=0$ or
\begin{eqnarray}
\widetilde{m}^2=-i\eta.
\end{eqnarray}
As $\eta \rightarrow 0$, the second term in  equation (\ref{i2_eqmass})
is exponentially small compared to the first term and, 
therefore, it can be neglected.  Thus,  the leading contribution 
to the integral $I_2^{(d)}(0,0; q^2 )$ reads
\begin{eqnarray}
&&I_2^{(d)}(0,0; q^2 )=\frac{1}{i \pi^{d/2}} 
\int \frac{d^dk_1}{[(k_1-q)^2+i\eta]
[k_1^2+i\eta]}
\nonumber \\
&& \nonumber \\
&&~~~~~~~~~~~~~~~~~~=
\frac{-\pi^{\frac32}~(-q^2-4i \eta )^{\frac{d}{2}-2}}
{2^{d-3}  \Gamma\left(\frac{d-1}{2}\right)
 \sin \frac{\pi d }{2}} +
 O({\rm max}(|\eta|, |\eta|^{(d-2)/2})).
\label{I2_mmq}
\end{eqnarray}
We simplified the first term in (\ref{i2_eqmass}) by dropping terms
proportional to  $\eta$, keeping the structure of the branch point
in the vicinity of $q^2=0$. 
The leading term in (\ref{I2_mmq}) is in agreement with  (\ref{xi2_00q}).

%%%%%%%%%%%%%%%%%%%%%%%%%%%%%%%%%%%%%%%%%
\section{Appendix C}
%%%%%%%%%%%%%%%%%%%%%%%%%%%%%%%%%%%%%%%%%
For the sake of completeness, we  present  in this appendix useful 
formulae  for the Appell $F_1$ and Gauss $_2F_1$ hypergeometric functions.

\subsection{ The $_2F_1$ Gauss hypergeometric function }

a) Series representation:
\begin{eqnarray}
\label{eq:f21_def}
{_2F_1}\left(\alpha, \beta, \gamma,x\right) 
&=& \sum_{m=0}^{\infty} \frac{(\alpha)_m (\beta)_m
}{(\gamma)_m} \,\frac{x^m}{m!} 
\end{eqnarray}
b) Integral representation:
\begin{eqnarray}
&&\Fh21\Fx{\alpha,\beta}{\gamma} =
\frac{\Gamma(\gamma)}{\Gamma(\beta)\Gamma(\gamma-\beta)}
~~ \int_0^1 du  \, u^{\beta-1}
(1-u)^{\gamma-\beta-1} (1-u x)^{-\alpha},
\nonumber \\
&& \nonumber \\
&&
~~~~~~~~~~~~~~~~~~~~~~~~~~~~~~~~~~~~~~~~~~~~~~~
{\rm Re}(\beta)>0, \quad {\rm Re}(\gamma-\beta) >0.
\end{eqnarray}

\subsection{The $F_1$ Appell function}

a) Series representation 
\begin{eqnarray}
F_1(\alpha,\beta, \beta',\gamma;~x, y)
=\sum_{n=0}^{\infty} 
\frac{(\alpha)_n~(\beta)_n}
     {(\gamma)_n} \frac{x^n}{n!}
~\Fh21\Fy{\alpha+n,\beta'}{\gamma+n}.
\end{eqnarray}

\begin{equation}
F_1(\alpha,\beta,\beta';~\gamma;~x,y)=\sum_{n=0}^{\infty}
\sum_{m=0}^{\infty}\frac{(\alpha)_{n+m}(\beta)_{n}
(\beta')_m}
{(\gamma)_{n+m}~n!m!}~x^ny^m,
\end{equation}
b) Integral representation 
\begin{equation}
F_1(\alpha,\beta,\beta';~\gamma;~x,y)=
\frac{\Gamma(\gamma)}{\Gamma(\alpha)\Gamma(\gamma-\alpha)}
\int_0^1\frac{u^{\alpha-1}(1-u)^{\gamma-\alpha-1}}
{(1-ux)^{\beta}(1-uy)^{\beta'}}~du
\end{equation}
c) Analytic continuation for the Appell function $F_1$ at large
argument $x$  \cite{OlssonJMP5}, \cite{Bezrodnykh2017}:
\begin{eqnarray}
&&F_1(\alpha,\beta,\beta',\gamma;x,y)
\nonumber \\
&&
=\frac{\Gamma(\gamma)\Gamma(\beta-\alpha)}
{\Gamma(\beta)\Gamma(\gamma-\alpha)}
(-x)^{-\alpha}
F_1\left(\alpha,1+\alpha-\gamma,\beta',
1+\alpha-\beta,\frac{1}{x},\frac{y}{x}
\right)
\nonumber \\
&&+\frac{\Gamma(\gamma)\Gamma(\alpha-\beta)}
{\Gamma(\alpha)\Gamma(\gamma-\beta)}
(-x)^{-\beta}~G_2\left(\beta,\beta',\alpha-\beta,
1+\beta-\gamma;-\frac{1}{x},-y\right),
\end{eqnarray}
where
\begin{equation}
G_2(a_1,a_2,b_1,b_2;x,y)=\sum_{m=0}^{\infty}
\sum_{n=0}^{\infty}(a_1)_m(a_2)_n
(b_1)_{n-m}(b_2)_{m-n}\frac{x^m}{m!}
\frac{y^n}{n!},~~~~~|x|<1,~~|y|<1.
\end{equation}

\providecommand{\href}[2]{#2}
%\bibliography{funcred.bib}

\end{document}